\documentclass[a4paper,fleqn]{cas-sc}

\usepackage[section]{placeins} 
\usepackage{amssymb}
\usepackage{amsthm}
\usepackage{amsmath}
\usepackage{upgreek}
\usepackage{pdflscape}
\usepackage{listings}
\usepackage{multirow}
\usepackage[percent]{overpic}
\usepackage{multicol}
\usepackage{color}
\usepackage{mathtools}
\usepackage{stmaryrd}
\usepackage{xfrac}
\usepackage[capitalise]{cleveref}
\usepackage{booktabs}
\usepackage[section]{placeins} 
\usepackage{calc}
\usepackage[titletoc]{appendix}
\usepackage[binary-units=true]{siunitx}
\usepackage[sort&compress,square,numbers]{natbib}

\usepackage{graphicx}
\usepackage{graphics}
\usepackage{inconsolata}
\usepackage{wrapfig}
\usepackage{float}
\usepackage{subfig}
\usepackage[percent]{overpic}
\usepackage{varwidth}
\usepackage{tikz}
\usepackage{pgfplots}
\usepackage{adjustbox}
\usetikzlibrary{arrows,automata,backgrounds,fit,intersections,matrix,mindmap,petri,positioning,shapes,snakes}
\usepackage{tikz-layers}
\usepgfplotslibrary{fillbetween}
\pgfplotsset{compat=newest}
\usepgfplotslibrary{colorbrewer}
\usepgfplotslibrary{patchplots}
\usepgfplotslibrary[colorbrewer]
\usetikzlibrary{pgfplots.colorbrewer}
\usetikzlibrary[pgfplots.colorbrewer]
\usetikzlibrary{patterns}
\usepgfplotslibrary{units}
\usetikzlibrary{spy}
\usepackage{pgfplotstable}
\usepackage{arrayjobx}
\graphicspath{ {./figs/} }
\usetikzlibrary{external}

\newlength\myheight
\newlength\mydepth
\settototalheight\myheight{Xygp}
\newcommand*\inlinegraphics[1]{%
  \settototalheight\myheight{Xygp}%
  \settodepth\mydepth{Xygp}%
  \raisebox{-\mydepth}{\includegraphics[height=\myheight]{#1}}%
}
\newcommand\orcid[1]{\href{https://orcid.org/#1}{\inlinegraphics{orcid_16x16.png}}}

\makeatletter
\def\BState{\State\hskip-\ALG@thistlm}
\makeatother

\newdefinition{definition}{Definition}[section]

\newcommand\px[2]{\frac{\partial #1}{\partial {#2}}}
\newcommand\pxi[3]{\frac{\partial^{#1}#2}{\partial {#3}^{#1}}}
\newcommand\dx[2]{\frac{\mathrm{d} #1}{\mathrm{d} #2}}

\newcommand\pxvar[2]{\partial_{#2} #1}

\newcommand{\half}{{\frac{1}{2}}}

\newcommand\avg[1]{\langle {#1}\rangle}

\begin{document}

\title[mode=title]{Artificial Compressibility Approaches in Flux Reconstruction for Incompressible Viscous Flow Simulations}
\shorttitle{Artificial Compressibility and FR}
\shortauthors{Trojak et al.}

\author[1]{W. Trojak}[orcid=0000-0002-4407-8956]
\cormark[1]
\ead{wtrojak@ic.ac.uk}
\cortext[cor1]{Corresponding author}

\address[1]{Department of Aeronautics, Imperial College London, London, SW7 2AZ}
\address[2]{Department of Aerospace Engineering, IIT Madras, Chennai, 600036}
\address[3]{Department of Mechanical and Aerospace Engineering, Brunel University, Uxbridge, UB8 3PH}
\address[4]{Department of Ocean Engineering, Texas A\&M University, College Station, TX 77843}
\address[5]{Department of Aerospace Engineering, Texas A\&M University, College Station, TX 77843}

\author[2]{ N.R. Vadlamani}[]
\ead{nrv@smail.iitm.ac.in}

\author[3]{ J. Tyacke}[]
\ead{James.Tyacke@brunel.ac.uk}

\author[4]{ F.D. Witherden}[]
\author[5]{ A. Jameson}[]

\begin{abstract}
    Several competing artificial compressibility methods for the incompressible flow equations are examined using the high-order flux reconstruction method. The established artificial compressibility method (ACM) of \citet{Chorin1967} is compared to the alternative entropically damped (EDAC) method  of \citet{Clausen2013}, as well as an ACM formulation with hyperbolised diffusion. While the former requires the solution to be converged to a divergence free state at each physical time step through pseudo iterations, the latter can be applied explicitly. We examine the sensitivity of both methods to the parameterisation for a series of test cases over a range of Reynolds numbers. As the compressibility is reduced, EDAC is found to give linear improvements in divergence whereas ACM yields diminishing returns. For the Taylor--Green vortex, EDAC is found to perform well; however on the more challenging circular cylinder at $Re=3900$, EDAC gives rise to early transition of the free shear-layer and over-production of the turbulence kinetic energy. This is attributed to the spatial pressure fluctuations of the method. Similar behaviour is observed for an aerofoil at $Re=60,000$ with an attached transitional boundary layer. It is concluded that hyperbolic diffusion of ACM can be beneficial but at the cost of case setup time, and EDAC can be an efficient method for incompressible flow. However, care must be taken as pressure fluctuations can have a significant impact on physics and the remedy causes the governing equation to become overly stiff.
\end{abstract}

\begin{highlights}
\item Three artificial compressibility methods are compared in a high-order framework.
\item Entropically damped artificial compressibility (EDAC) can be an effective alternative for unsteady incompressible simulations.
\item Spatial pressure waves in EDAC are found to be responsible for incorrect transition, separation, and reattachment dynamics.
\end{highlights}




\maketitle

\section{Introduction}\label{sec:intro}
    There is a growing desire across many industries to gain detailed insight into transient flow phenomena. In many cases, however, this is difficult to achieve with experimentation due to high costs and limited access to observe key physics. Over the recent decades, there has been growing interest in high-order methods in engineering applications as, compared to low order approaches, the resolution and computational efficiencies achievable can make many intractable problems tractable~\citep{Vincent2016}. A further factor in this success has a shifting desire from modelled simulations, of which Reynolds averaged Navier--Stokes (RANS) may be considered representative, to the scale-resolving methods such as large eddy simulation (LES). Although RANS has seen great success in several industrially relevant problems, such as transonic wing shape optimisation~\citep{Kim2002} and stall prediction in transonic fans~\citep{Williams2020}. The approximations of the model limit its accuracy for various problems involving transition, strong streamline curvature, and relaminarization. Durbin~\citep{Durbin2018} discusses recent advances, remaining and new challenges to RANS turbulence modelling as application areas widen. Scale resolving simulations, on the other hand, attempt to directly resolve the majority of the physical length scales on the grid. Such approaches can elucidate complex flow physics, especially those rooted in the small scale motions. In explicit LES methods, the effect of unresolved length scales are modelled using a sub-grid scale (SGS) model, often coupled to a filter. An alternative paradigm that has gained wider adoption, is implicit LES (ILES) where no SGS model is used and dissipation of the small scales is handled implicitly by the numerical dissipation of the discretisation. The spatial scheme we will use to perform ILES is the high-order flux reconstruction method of \citet{Huynh2007}, in the PyFR implementation~\citep{Witherden2014}.
    
    A challenge occurs in the limit as Mach number, $M$, tends to zero. The compressible flow equations become increasingly stiff, ultimately resulting in an elliptic equation for the pressure field at $M=0$. If the compressible Navier--Stokes equations are used in this low Mach limit, the stiffness can lead to prohibitively expensive calculations. In the case of the method of \citet{Roe1981} it also leads to excessive numerical dissipation. These problems can be alleviated by the use of a low Mach preconditioners, such as that of \citet{Turkel1987}. Alternatively, in the incompressible regime a separate solver is used to solve the pressure Poisson equation. However, it is non-trivial to produce a scalable solver for the Poisson equation, although new methods such as that of \citet{Fortunato2019} are beginning to confront aspects of this. Furthermore, within finite element methods ensuring solution compatibility requires detailed analysis that is dependent on order and element type, among other factors~\citep{Elman2014}. 
    
    A single solver is preferable to make use of established tools and optimisations, and to simplify work flows. Artificial compressibility approaches allows the use of established compressible tool to calculated state solutions of incompressible flows. The first such method was the artificial compressibility method (ACM) of \citet{Chorin1967}, which can be interpreted as assuming constant entropy together with an artificial compressibility to relax the pressure and velocity field onto a divergence free solution. ACM was extended to the calculation of unsteady flows by \citet{Rogers1991} by using it as a method to solve the implicit equations for each physical time step.
    Later \citet{Jameson1991} interpreted the relaxation as a pseudo-time dimension, using explicit time stepping to perform the relaxation. More recently, \citet{Nishikawa2007} proposed a general technique where diffusion terms are hyperbolised. This has the advantage of stability scaling with $h^{-1}$ --- rather than $h^{-2}$ --- for some mesh spacing $h$. This approach fits naturally with ACM and has previously been investigated by \citet{Ahn2020} and \citet{Trojak2021}, who introduced novel techniques to optimise the computational implementation of ACM with hyperbolised diffusion (ACM-HD).
    
    A major issue with ACM is the requirement to converge the pressure and velocity field for each time step, which can be costly, although some convergence acceleration methods have proved successful~\citep{Loppi2019,Loppi2021}. An alternative is the entropically damped artificial compressibility (EDAC) method introduced by \citet{Clausen2013}. Here entropy is not fixed, but density fluctuations are minimised. This results in a similar system of equations as ACM, with the primary difference being a pressure diffusion term. This approximation leads to  time dependent equations which produce an almost divergence free flows thus enabling explicit time stepping to be used. Further studies have shown the method is effective on both model problems and turbulent channel flows~\citep{Achu2021,Toutant2018,Dupuy2020}.
    
    These three methods, ACM, ACM-HD, and EDAC, have benefits and drawbacks. We examine there relative costs and effectiveness for relevant unsteady turbulent LES. To the authors knowledge the EDAC system has not previously been used with FR and we wish to understand its effectiveness and the effect of varying the compressibility parameter within a high-order approach. Furthermore, as the EDAC system will be applied as a conservative equation, we wish to further the understanding of the Riemann problem as it forms an important part of the FR method~\citep{Moura2017}. We also wish to better understand the effectiveness of the ACM approaches and how the runtime of the various methods compare on GPU hardware.
    
    To this end, in \cref{sec:prelim} we introduce the high-order FR approach used in this work. In \cref{sec:govern}, we detail the systems of governing equation for incompressible flow and the artificial compressibility approaches studied in this work. We also explore aspects of the eigen-structure of these equations. Some additional details on the Riemann problem are included in \cref{sec:riemann_acmhd,sec:riemann}. Then, in \cref{sec:results}, the main numerical results on unsteady turbulent test cases are presented. Finally, conclusions are drawn in \cref{sec:conclusions}.

\section{Preliminaries}\label{sec:prelim}
    \noindent
    In this work we consider artificial compressibility approaches solved via the high-order method flux reconstruction (FR)~\citep{Huynh2007}, as implemented in the PyFR tool~\citep{Witherden2014}. The original FR method of \citet{Huynh2007} has been adapted to handle problems including advection-diffusion equations on element typologies such as simplicies, hypercubes, prisms, and affine pyramids. For completeness the FR method is summarised here, where for brevity we restrict the statement of the method to one dimension. \citep{Williams2013,Witherden2015} and references therein are recommended for applications to alternative topologies.
    
    Characteristic of finite element methods, FR uses a partition of the domain $K$ into $N$ conformal sub-domains such that $K=\cup_{i=1}^NK_i$ and $K_i\cap K_j=\emptyset$ if $i\neq j$. Within each sub-domain, a number of nodes are positioned such that a Lagrange finite element can be formed for a conservation equation of the form 
    \begin{equation}
        \px{u}{t} + \px{f}{x} = 0, \quad \mathrm{for} \quad u(x,t):K\times\mathbb{R}_+\mapsto U\subset \mathbb{R} \quad \mathrm{and} \quad f:U\mapsto \mathbb{R}, \quad \mathrm{with} \quad u(x,0)=u_0(x),
    \end{equation}
    where in this description of FR we will assume $K$ is periodic. It is typical to use a reference domain $\hat{K}$ with the projection $T_i:K_i\mapsto\hat{K}$ as this makes all the operators the same for a given element topology which has some clear computational benefits. For a line, quadrilateral, or hexahedral element a common choice of reference domain is: $\hat{K}=[-1,1]$, $\hat{K}=[-1,1]^2$, or $\hat{K}=[-1,1]^3$ respectively.
    
    For the sub-domain $K_i$ and points $x_{ij}\in K_i$, a Lagrange finite element can be formed giving the solution and flux polynomials as:
    \begin{equation}\label{eq:ns_poly}
        \hat{u}^\delta_i = \sum^{n_s}_{j=1}u_{i}(x_{ij})l_j(r) \quad \mathrm{and} \quad \hat{f}^\delta_i = \sum^{n_s}_{j=1}f(u_{i}(x_{ij}))l_j(r), \quad \mathrm{for} \quad r\in\hat{K}, 
    \end{equation}
    where $n_s$ is the number of nodal value within each element, and $l_j$ is the j-th Lagrange polynomial defined as
    \begin{equation}
        l_j(\xi) = \prod^{n_s}_{\substack{k=1 \\ k\neq j}}\frac{r - r_k}{r_j - r_k},
    \end{equation}
    for the set of reference points $\{r_1, \dots, r_{n_s}\}\in \hat{K}$. In \cref{eq:ns_poly}, $\hat{\bullet}$ symbolises this is in the reference domain and $\delta$ indicates this corresponds to a piece-wise discontinuous approximation. The flux reconstruction algorithm provides a method to calculate the gradient of the flux function, $f$, corresponding to a $\mathcal{C}^0$ approximation of $f$ in $K$. This gradient approximation is given as
    \begin{equation}
        \px{f_i}{x}\approx \left(\dx{T_i}{r}\right)^{-1}\left(\dx{\hat{f}^\delta}{r} + \left(f^I_L - f^\delta_L\right)\dx{h_L}{r} + \left(f^I_R - f^\delta_R\right)\dx{h_R}{r} \right),
    \end{equation}
    where $\bullet_L$ and $\bullet_R$ are the projection to the left and right interfaces. Then $f^I_L$ is the common interface value formed by using $u^\delta_{i,L}$ and $u^\delta_{i-1,R}$ and similarly for $f^I_R$. Some degree of upwinding should be applied when calculating the common inviscid interface flux in order to stabilise the method. This may be provided from one of a number of approximate Riemann solvers~\citep{Toro2009}. These common interface values are applied to the element via polynomial correction functions $h_L(r)$ and $h_R(r)$. To enforce the common value, these functions have the conditions that $h_L(-1)=h_R(1)=1$ and $h_L(1)=h_R(-1)=0$. 
    
    Once the approximate flux divergence is calculated, one of a number of ODE integration techniques can be used. In this work, Runge--Kutta (RK) time integration will be used for explicit time stepping. When solving the dual time systems that arise in the ACM system, it is logical to use an implicit time scheme. In this case, the BDF2 method is used, coupled to an explicit RK smoother in pseudo-time. Throughout this work, we will use the adaptive low storage RK procedure of \citet{Kennedy2000}, particularly the RK3(2)4[2R+] method. This is a third order, four stage method where the embedded scheme is second order. The embedded system is used to predict the error~\citep{Hairer1993} and a PI controller can be used to set either a global or local time step~\citep{Witherden2015,Loppi2019} for the physical time step or pseudo-time step, respectively.
\section{Governing Equations}\label{sec:govern}

\subsection{Incompressible flow}
\noindent
    The incompressible Navier--Stokes equations for a single phase and constant density can be written as
    \begin{equation}
        \px{\mathbf{V}}{t} + \mathbf{V}\cdot\nabla\mathbf{V} + \nabla P = \nu\nabla^2\mathbf{V} \quad \mathrm{and}\quad \nabla\cdot\mathbf{V} = 0,
    \end{equation}
    where $\mathbf{V}=[u,v,w]^T$ is the velocity vector, $P$ is the pressure, and $\nu$ is the kinematic viscosity. Taking the divergence of this equation and enforcing a solenoidal velocity field gives the pressure Poisson equation
    \begin{equation}
        \nabla^2P = -\nabla\cdot(\mathbf{V}\cdot\nabla\mathbf{V}).
    \end{equation}
    This equation gives a closed form for pressure up to a constant offset. The difficulty arises from the global nature of solutions to the Poisson equation and the complexities of solving this in parallel with the same efficiency as hyperbolic and parabolic equations.
    
\subsection{Artificial Compressibility}\label{ssec:acm}
\noindent
    The artificial compressibility method (ACM) introduced by \citet{Chorin1967} can be derived by starting from the compressible Navier--Stokes equations and imposing constant entropy to close an incompressible model~\citep{Clausen2013}. A pseudo-time derivative can be introduced, which interprets a physical time step as converging the field variables in pseudo-time to a steady state. This ACM system of equations can be written in conservative form as: 
    \begin{equation}
        \px{}{t}\begin{bmatrix}
        0 \\ u \\ v \\ w
        \end{bmatrix} + \px{}{\tau}\begin{bmatrix}
        P \\ u \\ v \\ w
        \end{bmatrix} + \nabla\cdot\begin{bmatrix}
            \zeta u & \zeta v & \zeta w \\ u^2 + P & uv & uw \\ uv & v^2 + P & vw \\ uw & vw & w^2 + P
        \end{bmatrix} = \nu\begin{bmatrix}
            0 & 0 & 0 \\
            \pxvar{u}{xx} & \pxvar{u}{yy} & \pxvar{u}{zz} \\
            \pxvar{v}{xx} & \pxvar{v}{yy} & \pxvar{v}{zz} \\
            \pxvar{w}{xx} & \pxvar{w}{yy} & \pxvar{w}{zz} \\
        \end{bmatrix},
    \end{equation}
    where $\tau$ is the pseudo-time variable, and $t$ is the physical time. Here, $\zeta$ is a measure of the compressibility of the flow in pseudo-time, defining an artificial Mach number of the flow as $\zeta=1/M^2$, which leads to velocity divergence being balanced by pressure gradients in the pseudo-time dimension. This equation shows the key behaviour of incompressible flow, as $\partial_\tau P \to 0$ the velocity field converges on to a divergence free solution $\nabla\cdot\mathbf{V}\to0$. Importantly though there is no explicit elliptic equation to solve, and a solver developed for conservation equations may be readily applied to ACM.
    
    To understand the prorogation of information in the system further, consider the flux vector in $x$ is defined as $\mathbf{f}=[\zeta u, u^2 +P, uv, uw]^T$, then the inviscid flux Jacobian can be found to be:
    \begin{equation}
        \mathbf{A}_x =  \px{\mathbf{f}}{\mathbf{U}} = \begin{bmatrix}
            0 & \zeta & 0 & 0 \\
            1 & 2u & 0 & 0 \\
            0 & v & u & 0\\
            0 & w & 0 & u
        \end{bmatrix},
    \end{equation}
    where $\mathbf{U}=[P,u,v,w]^T$. This is used to highlight the difference between the conserved variables in pseudo-time and real-time. The eigenvalue of this Jacobian can be found to be:
    \begin{equation}
        \lambda_1 = \lambda_2 = u, \quad \lambda_3 = u - a, \quad \mathrm{and} \quad \lambda_4 = u + a,
    \end{equation}
    for $a^2 = u^2 + \zeta$. The repeated eigenvalue indicates that the inviscid system is linearly degenerate, and through calculation of the eigenvectors, it can be seen that this results in a contact discontinuity in the Riemann problem with discontinuities in the tangential velocity components, $v$ and $w$. As the contact is not stationary, exact or structure approximating Riemann solvers can be developed without complication via approaches such as that of \citet{Elsworth1992}. However, in this work only Rusanov approximate Riemann solvers were applied to give a common interface flux via:
    \begin{equation}
        \mathbf{f}^I = \frac{1}{2}\left(\mathbf{f}_L + \mathbf{f}_R\right) - \frac{1}{2}S_\mathrm{max}\left(\mathbf{U}_R - \mathbf{U}_L\right),
    \end{equation}
    where the Davis type maximum wavespeed estimate is used as:
    \begin{equation}
        S_\mathrm{max} = \max{\left(|u_L|+a_L, |u_R|+a_R\right)}.
    \end{equation}
    Importantly this shows that the maximum absolute eigenvalue scales with $\sqrt{\zeta}$, ie the stiffness will scale with one over the fictional Mach number. 

\subsection{Artificial Compressibility - Hyperbolised Diffusion}\label{ssec:acmhd}
    \noindent
    The hyperbolic diffusion method of \citet{Nishikawa2007} aims to remove parabolic terms in governing equations through additional auxiliary equations which, once converged, yield the gradients of the conserved variables. As an example consider the linear advection-diffusion equation:
    \begin{equation*}
        \px{\phi}{t} + \px{\phi}{x} = \nu\pxi{2}{\phi}{x},
    \end{equation*}
    the diffusion term can be hyperbolised by adding an auxiliary equation to give the system:
    \begin{subequations}
        \begin{align*}
            \px{\phi}{t} + \px{\phi}{\tau} + \px{}{x}(\phi - \nu\psi) &= 0, \\
            \px{\psi}{\tau} + \px{}{x}\left(-\frac{\phi}{T}\right) &= -\frac{\psi}{T}.
        \end{align*}
    \end{subequations}
    Here pseudo-time is again used to converge the system, where $T$ is a preconditioning parameter to account for the differing stiffness of the equations. This system can be understood by considering $\partial_\tau\psi\to 0$, as this happens $\psi\to\partial_x\phi$.

    As the hyperbolic diffusion method and the artificial compressibility method can both be formulated to use pseudo-transient continuation, ACM is a good candidate for hyperbolic diffusion. The new ACM-HD governing conservation equations can then be expressed in three dimensions as:

    \begin{equation}\label{eq:acmhd_nd}
            \px{}{t}\begin{bmatrix}
                0 \\ \mathbf{V} \\ \mathbf{0} \\ \mathbf{0} \\ \mathbf{0} 
            \end{bmatrix} 
            + \px{}{\tau}\begin{bmatrix}
                P \\ \mathbf{V} \\ \mathbf{q} \\ \mathbf{r} \\ \mathbf{s}
            \end{bmatrix} 
            + \nabla\cdot\begin{bmatrix}
                \zeta \mathbf{V}^T \\ \mathbf{V}\otimes\mathbf{V} + P\mathbf{I} - \nu \mathbf{S}^T \\ -\frac{1}{T}\mathbf{V}\otimes\mathbf{I}
            \end{bmatrix} = 
            -\frac{1}{T}\begin{bmatrix}
                0 \\ \mathbf{0} \\ \mathbf{q} \\ \mathbf{r} \\ \mathbf{s} 
            \end{bmatrix},
    \end{equation}
    where $\mathbf{V}=[u,v,w]^T$. Additionally we have the vectors $\mathbf{q}$, $\mathbf{r}$, and $\mathbf{s}$, and the matrix $\mathbf{S}=[\mathbf{q},\mathbf{r},\mathbf{s}]$. These vectors will form the gradient of $u$, $v$, and $w$, respectively such that once converged $\mathbf{q}\to[\pxvar{u}{x}, \pxvar{u}{y}, \dots]^T$, $\mathbf{r} \to [\pxvar{v}{x},\dots]^T$, and $\mathbf{s} \to [\pxvar{w}{x},\dots]^T$.
    
    The resulting Jacobian matrices, to the authors knowledge, have not been well analysed in the literature. To see the effect of hyperbolising diffusion on the stiffness consider the Jacobian of the invisicid flux in the $x$ direction:
    \begin{equation}
        \mathbf{A}_x = \px{\mathbf{f}}{\mathbf{U}} = \begin{bmatrix}
            0 & \zeta & 0 & 0 & 0 & 0 & 0 & 0 & 0 & 0 & 0 & 0 & 0\\
            1 & 2u & 0 & 0 & -\nu & 0 & 0 & 0 & 0 & 0 & 0 & 0 & 0\\
            0 & v & u & 0 & 0 & 0 & 0 & -\nu & 0 & 0 & 0 & 0 & 0 \\
            0 & w & 0 & u & 0 & 0 & 0 & 0 & 0 & 0 & -\nu & 0 & 0 \\
            0 & -1/T & 0 & 0 & & & & & & & &\\
            0 & 0 & 0 & 0 & & & & & & & & & \\
            0 & 0 & 0 & 0 & & & & & & & & & \\
            0 & 0 & -1/T & 0 & & & & & & & & \\
            0 & 0 & 0 & 0 & & & & & \mathbf{0} & & & & \\
            0 & 0 & 0 & 0 & & & & & & & & &\\
            0 & 0 & 0 & -1/T & & & & & & & & \\
            0 & 0 & 0 & 0 & & & & & & & & & \\
            0 & 0 & 0 & 0 & & & & & & & & &\\
        \end{bmatrix}.
    \end{equation}
    This yields the eigenvalues:
    \begin{equation}
        \lambda_1= \dots = \lambda_7 = 0, \quad \lambda_8=\lambda_9 = \frac{u}{2} - c, \quad \lambda_{10}=\lambda_{11}= \frac{u}{2} + c, \quad \lambda_{12} = u - b, \quad \lambda_{13}= u + b,
    \end{equation}
    where:
    \begin{equation}
        b^2 = u^2 + \zeta + \frac{\nu}{T}, \quad \mathrm{and}\quad c^2=\frac{u^2}{4} + \frac{\nu}{T}.
    \end{equation}
    Consequently, a Davis estimate of the maximum absolute eigenvalue can be taken as:
    \begin{equation}
        S_\mathrm{max} = \max{(|u_L|+b_L, |u_R|+b_R)}.
    \end{equation}
    Consideration must be given as to how the preconditioning parameter, $T$, is set. If $T=\mathcal{O}(\nu)$ is used then the stiffness will be approximately independent of the viscosity, a similar argument has been proposed by \citet{Nishikawa2018}. The repeated eigenvalues also highlight another feature of this governing equation, namely that there will be a linear degeneracy in the Riemann problem, resulting in a stationary contact discontinuity. 
    This introduces some difficulties in using an exact Riemann solver or an HLLC type approximate solver to produce the common interface flux, as a correctly resolved problem would have a contact discontinuity at the interface. The alternative is to again use a Rusanov type Riemann approximate solver with Davis wavespeed estimate for $[P,u,v,w]^T$. Central differencing was used for the remaining terms owing to the diffusive nature of the auxiliary equations. More details of the Riemann problem relating to ACM-HD are given in \cref{sec:riemann_acmhd}.
    
    An advantage of this system should be apparent, we solely have first order derivatives in the system and consequently the numerical stability is only dependent on the advection scheme, which scales with $h^{-1}$. This is at the cost of additional equations, but this can be offset by only requiring the simpler advection FR algorithm, as well as an increased rate of convergence reported for hyperbolic diffusion~\citep{Nishikawa2018}.
    
\subsection{Entropically Damped Artificial Compressibility}\label{ssec:edac}
\noindent
    \citet{Clausen2013} introduced the entropically damped artificial compressibility (EDAC) method where closure was achieved by minimising density variations, rater than constant entropy as in ACM. This leads to the following evolution equation for the pressure
    \begin{equation}
        \px{P}{t} + \mathbf{V}\cdot\nabla P + \frac{1}{M^2}\nabla\cdot\mathbf{V} = \frac{1}{Re}\nabla^2P,
    \end{equation}
    where $Re$ is a Reynolds number. The key difference is that parabolic regularisation introduced to the mass equation removes the need for pseudo-transient continuation, ie  explicit time stepping can be used. The name EDAC then stems from the relationship between pressure diffusion and entropy. If entropy is defined via the functional $\sigma=\log(P)$, then pressure diffusion terms will provide damping to the entropy field.  To see this consider the companion entropy equation which may be calculated explicitly, following the general form of \citet{Dafermos2005}, for the entropy-flux pair $(\sigma, \mathbf{\Sigma})= (\log(P), \mathbf{V}\sigma)$:
    \begin{equation}
        \px{\sigma}{t} + \nabla\cdot\mathbf{\Sigma} = \frac{1}{Re P}\nabla^2P.
    \end{equation}
    Not only does this show how pressure diffusion effects the entropy, but it also shows that areas of low pressures will move the solution away from the the physical condition of constant entropy if accompanied by a second derivative of pressure. For example, the canonical case of laminar flow around a cylinder.
    
    In order to apply the EDAC method in the PyFR framework, it must be cast as a conservation law. If it is assumed that $\nabla\cdot\mathbf{V}=0$, then in three-dimensions we obtain
    \begin{equation}\label{eq:vars}
        \mathbf{Q} = \begin{bmatrix}
            P \\ u \\ v \\ w
        \end{bmatrix},
        \quad
        \mathbf{F}^\mathrm{inv} =  \begin{bmatrix}
            u(P+\zeta) & v(P+\zeta) & w(P+\zeta) \\ u^2 + P & uv & uw \\ uv & v^2 + P & vw \\ uw & vw & w^2 + P
        \end{bmatrix}, \quad \mathrm{and} \quad 
        \mathbf{F}^\mathrm{vis} =  \nu\begin{bmatrix}
            \pxvar{P}{x} & \pxvar{P}{y} & \pxvar{P}{z} \\
            \pxvar{u}{x} & \pxvar{u}{y} & \pxvar{u}{z} \\
            \pxvar{v}{x} & \pxvar{v}{y} & \pxvar{v}{z} \\
            \pxvar{w}{x} & \pxvar{w}{y} & \pxvar{w}{z} \\
        \end{bmatrix},
    \end{equation}
    for
    \begin{equation}\label{eq:edac_conv}
        \pxvar{\mathbf{Q}}{t} + \nabla\cdot\mathbf{F}^\mathrm{inv} = \nabla\cdot\mathbf{F}^\mathrm{vis},
    \end{equation}
    where $1/M^2 = \zeta$ and $1/Re = \nu$. This formulation shows a second aspect of EDAC, that pressure fluctuations are resolved spatially whereas in ACM they could be resolved in pseudo-time.
    
    To understand how the propagation of information differs in this system, we will again consider the inviscid flux Jacobian in $x$,  $\mathbf{f}=\mathbf{F}^\mathrm{inv}\cdot[1,0,0]^T$:
    \begin{equation}
        \px{\mathbf{f}}{\mathbf{Q}} = \begin{bmatrix}
            u & P+ \zeta & 0 & 0\\
            1 & 2u & 0 & 0\\
            0 & v & u & 0\\
            0 & w & 0 & u
        \end{bmatrix},
    \end{equation}
    then the eigenvalues can be found to be:
    \begin{equation}
        \lambda_1=\lambda_2 = u, \quad \lambda_{3} = \frac{3}{2}u - d, \quad \mathrm{and} \quad \lambda_{4} = \frac{3}{2}u + d,
    \end{equation}
    for $d^2 = u^2/4 + P + \zeta$. A Davis estimate may again be constructed as:
    \begin{equation}
        S_\mathrm{max} = \max{\left(\frac{3}{2}|u_L| + d_L, \frac{3}{2}|u_R| + d_R\right)}.
    \end{equation}
    Having defined these speeds, we will mainly use the Rusanov approximate Riemann solver throughout this work for EDAC. However, in \cref{sec:riemann} we explore the Riemann problem in more detail and define an HLLC approach which will also be investigated numerically. A further insight given from \cref{sec:riemann} is that the EDAC system cannot support a Riemann problem which leads to a solution containing two rarefaction waves. Although the concept of a shock in an artificial compressibility system is strange, this property can be thought of as consistent with the entropy dissipation of the method.
    
    Comparison of these eigenvalues to ACM shows a new pressure dependency which is interpreted as a consequence of resolving velocity divergence as pressure waves in space. Furthermore, assuming that $S_\mathrm{max}$ scales with the true maximum absolute eigenvalue, it can be seen that the EDAC system will often be significantly stiffer and, given that the permitted the use of explicit time stepping, this could lead to smaller stable time steps.
    

\section{Numerical Experiments}\label{sec:results}

\subsection{TGV at $Re=1600$}\label{ssec:tgv}
\begin{figure}[tbhp]
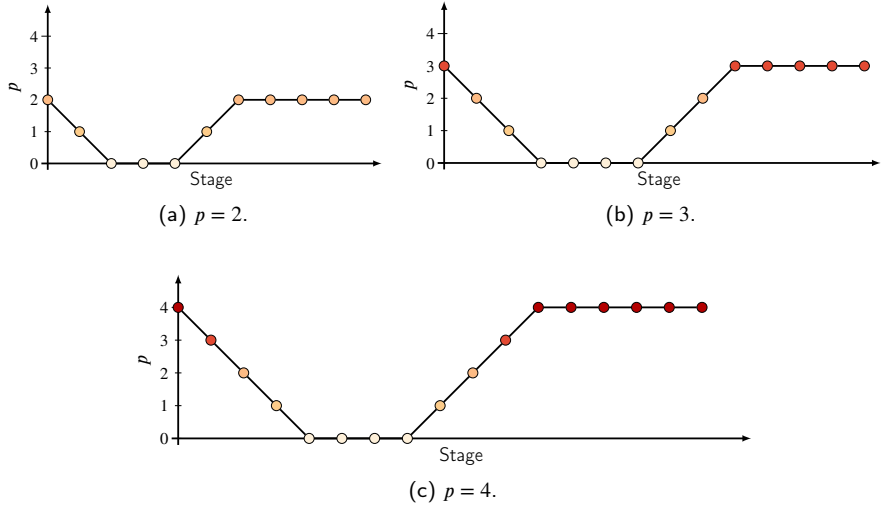

        \centering
        \subfloat[$p=2$.]{\adjustbox{width=0.317\linewidth,valign=b}{\input{figs/p_mg_cycle_p2}}}
        \subfloat[$p=3$.]{\adjustbox{width=0.4\linewidth,valign=b}{\input{figs/p_mg_cycle_p3}}}
        \\
        \subfloat[$p=4$.]{\adjustbox{width=0.51\linewidth,valign=b}{\input{figs/p_mg_cycle_p4}}}
        \caption{\label{fig:pmg}P-multigrid cycles used for various orders.}
    \end{figure}

    The Taylor--Green~\citep{Taylor1937} vortex is a well studied test case which exhibits three regimes: an inviscid laminar regime, turbulent transition through vortex sheet roll-up, and full homogeneous isotropic turbulent decay. The last regime is generally observed for $Re\gtrsim1000$~\citep{Brachet1983}. The initial condition is taken as:
    \begin{subequations}
        \begin{align}
            P &= \frac{1}{\gamma M^2} + \frac{1}{16}\cos{\left(2z + 2\right)}\left[\cos{(2x)} + \cos{(2y)}\right],\\
            u &= \sin{x}\cos{y}\cos{z},\\
            v &= -\cos{x}\sin{y}\cos{z},\\
            w &= 0,
        \end{align}
    \end{subequations}
    where a typical Mach number is $M=0.08$. When performing ACM-HD calculations, these terms are differentiated to give the initial gradients. The domain is a fully periodic cube $\Omega=[0,2\pi]^3$, which was partitioned into a mesh of regular hexahedral elements with approximately $128^3$ solution points depending on order. The solution and flux point locations in the reference domain were set using the tensor product of the Gauss--Legendre quadrature. For this case, unless otherwise stated, the common inviscid flux is calculated using a Rusanov type approach as detailed in \cref{sec:govern}. For the ACM and EDAC methods, the common viscous interface flux in calculated using an LDG approach with a small penalty, $\epsilon=0.1$. Two separate time integration schemes were used. For ACM and ACM-HD, BDF2 was coupled to an adaptive RK34 explicit smoother, details of which can be found in \cref{sec:prelim}. This same RK method was used for the explicit time stepping of the EDAC method; however, with global  adaptation of the explicit physical step rather than locally for the pseudo stepping. To further accelerate the convergence of the ACM and ACM-HD calculations, P-multigrid was used~\citep{Loppi2018}. The cycles used are shown in \cref{fig:pmg} as they were found to give rapid convergence and follow the suggested asymmetry of \citet{Trojak2020}. The exact configuration files can be found in the attached electronic supplementary material. 
    
    Several functionals were used to access the numerical performance of the AC methods. The first is enstrophy, defined as:
    \begin{equation}\label{eq:enst}
        \epsilon(t) = \frac{\nu}{|\Omega|}\int_\Omega \pmb{\omega}(t,\mathbf{x})\cdot\pmb{\omega}(t,\mathbf{x})\,\mathrm{d}\mathbf{x},
    \end{equation}
    where the vorticity is defined as $\pmb{\omega}=\nabla\times\mathbf{V}$. Reference DNS data for this functional was provided by Van Rees in private communication, containing results for a longer time period than that detailed in \citet{vanRees2011}. Two further functionals were examined to assess the quality of the incompressible solution:
    \begin{equation}
        s_s(t) = \frac{1}{|\Omega|}\int_\Omega\nabla\cdot\mathbf{V}(t,\mathbf{x})\,\mathrm{d}\mathbf{x} \quad \mathrm{and} \quad s_a(t) = \frac{1}{|\Omega|}\int_\Omega|\nabla\cdot\mathbf{V}(t,\mathbf{x})|\,\mathrm{d}\mathbf{x}.
    \end{equation}
    These are the volume averaged velocity divergence and absolute velocity divergence.
    
\begin{figure}[tbhp]
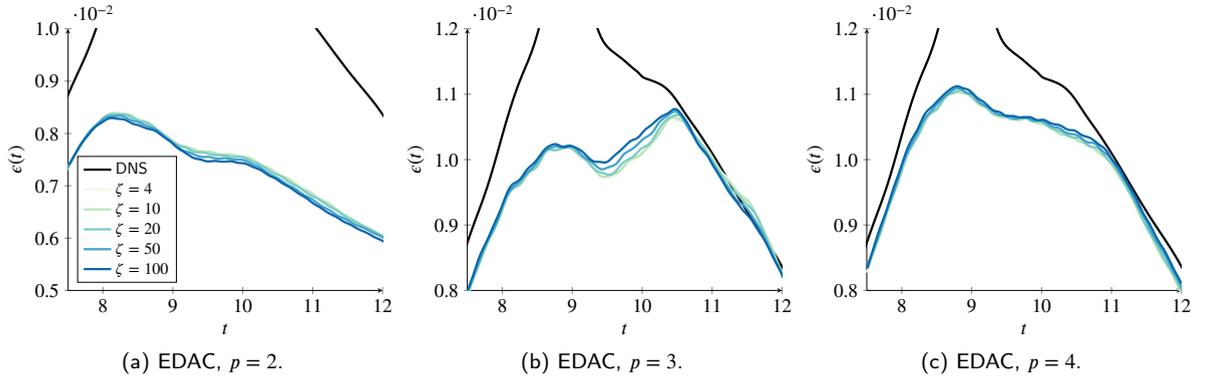

        \subfloat[EDAC, $p=2$.]{\adjustbox{width=0.32\linewidth, valign=b}{\input{figs/tgv_edac_p2_enstrophy}}}
        \subfloat[EDAC, $p=3$.]{\adjustbox{width=0.32\linewidth, valign=b}{\input{figs/tgv_edac_p3_enstrophy}}}
        \subfloat[EDAC, $p=4$.]{\adjustbox{width=0.32\linewidth, valign=b}{\input{figs/tgv_edac_p4_enstrophy}}}
        \caption{\label{fig:tgv_enst_edac}Taylor--Green vortex enstrophy comparison for $\mathrm{DoF}\approx128^3$. DNS data courtesy of van Rees, private communication.}
    \end{figure}

\begin{figure}[tbhp]
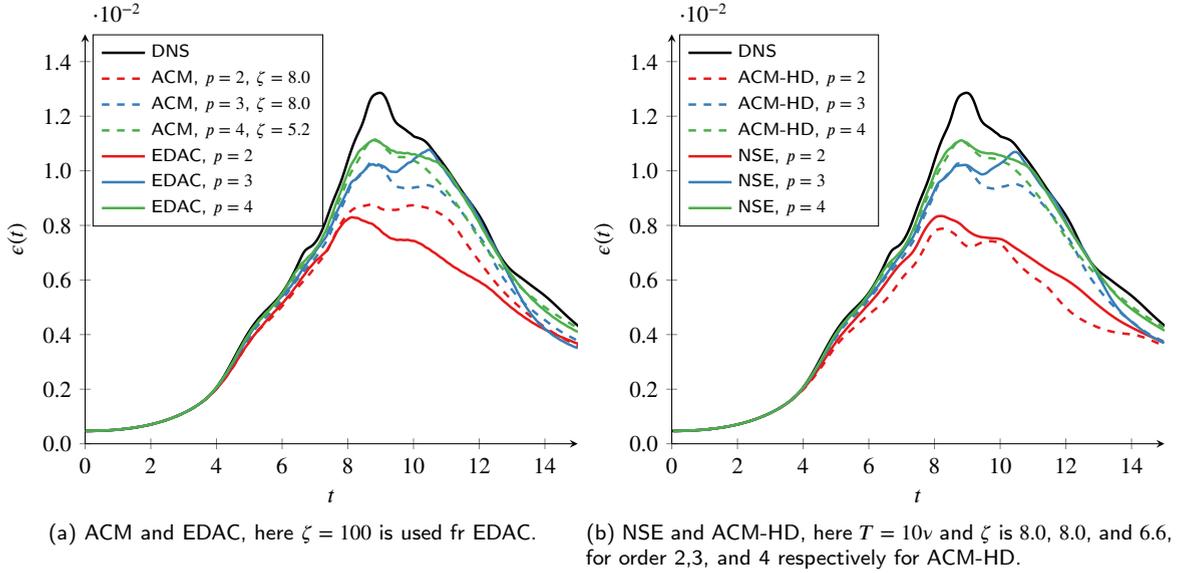

        \centering
        \subfloat[ACM and EDAC, here $\zeta=100$ is used fr EDAC.]{\label{fig:tgv_comp}\adjustbox{width=0.47\linewidth, valign=b}{\input{figs/tgv_comp}}}
        \subfloat[NSE and ACM-HD, here $T=10\nu$ and $\zeta$ is $8.0$, $8.0$, and $6.6$, for order 2,3, and 4 respectively for ACM-HD.]{\label{fig:tgv_comp_2}\adjustbox{width=0.47\linewidth, valign=b}{\input{figs/tgv_comp_2}}}
        \caption{\label{fig:tgv_enst}Comparison of TGV $\epsilon(t)$ with $\sim128^3$ DoF for various schemes and orders.}
    \end{figure}

    In order to evaluate the EDAC method, a sweep was performed for $\zeta\in\{3,\dots,100\}$ at orders 2, 3, and 4. Studying \cref{fig:tgv_enst_edac}, it can be observed that $\zeta$ does not have a significant effect on the enstrophy production for the EDAC method at this resolution, and the largest differences are seen at the peak dissipation, shown here. Comparison is made in \cref{fig:tgv_comp} between EDAC and the other ACM methods. This shows that after peak enstrophy production, there is a noticeable deviation between the results. A further comparison can be made from \cref{fig:tgv_enst} between EDAC and the compressible Navier--Stokes equations (NSE) at $M=0.08$. This makes it clear that EDAC and NSE are comparable, caused by both methods dissipating entropy due to compressibility and both systems having a similar Mach number. In contrast, the ACM and ACM-HD systems do not cause entropy dissipation and the results are evidently different from the entropically damped systems. There is also a notable disparity between the ACM and ACM-HD systems at $p=2$. The difference is due to \cref{eq:enst} being evaluated with the converged gradient terms in ACM-HD, which are one order higher than the reconstructed gradients used in standard ACM. This difference is more notable at lower $p$ as the polynomial space is more restricted.
    
    \begin{figure}[tbhp]
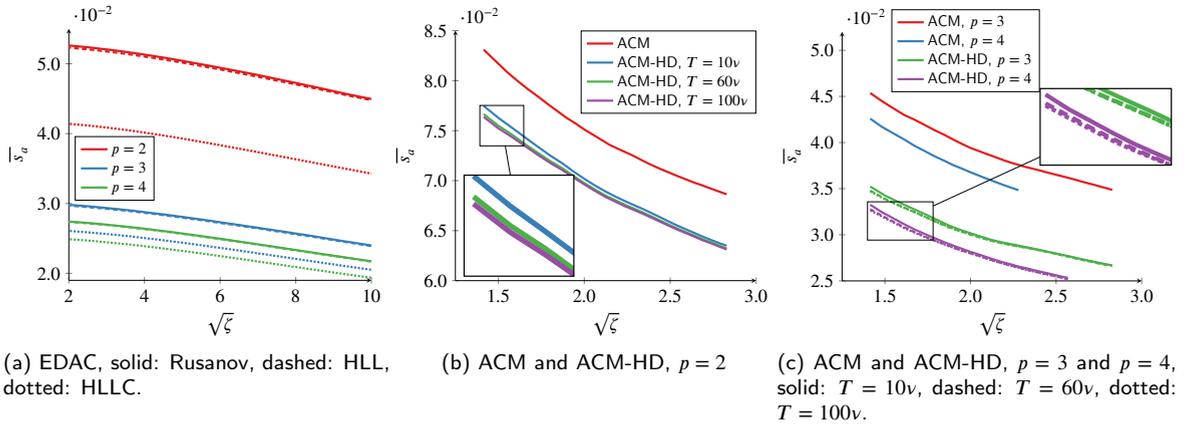

        \centering
        \subfloat[EDAC, solid: Rusanov, dashed: HLL, dotted: HLLC.]{\label{fig:tgv_edac_diva}\adjustbox{width=0.31\linewidth, valign=b}{\input{figs/tgv_diva_edac}}}
        \subfloat[ACM and ACM-HD, $p=2$]{\label{fig:tgv_diva_ac_p2}\adjustbox{width=0.31\linewidth, valign=b}{\input{figs/tgv_diva_ac_p2}}}
        \subfloat[ACM and ACM-HD, $p=3$ and $p=4$, solid: $T=10\nu$, dashed: $T=60\nu$, dotted: $T=100\nu$.]{\label{fig:tgv_diva_ac_p3p4}\adjustbox{width=0.32\linewidth, valign=b}{\input{figs/tgv_diva_ac_p3p4}}}
        \caption{\label{fig:tgv_diva}Variation of the TGV mean absolute divergence integral with $\sqrt{\zeta}$. Averaged over $t\in[0,15]$.}
    \end{figure}
    
    \cref{fig:tgv_diva} presents the time averaged values of $s_a$ against $\sqrt{\zeta}$. For EDAC, \cref{fig:tgv_edac_diva}, a linear relation between the time averaged $\overline{s_a}$ and $\sqrt{\zeta}$ is observed at high $\zeta$. Furthermore, by comparison to the ACM results, the divergence of EDAC is approximately $1.5\times$ lower. This linear relationship is not observed for ACM and ACM-HD. Instead, the benefit of increasing $\zeta$ is diminishing. For the dual time stepping of ACM and ACM-HD, a constant $\Delta t=0.01$ and locally adaptive $\Delta\tau$ with initial value of $0.0025$ was used for all runs. For this configuration at $p=4$, $\zeta>5.2$ was unstable with ACM and $\zeta>6.6$ for ACM-HD.  From the data presented in \cref{fig:tgv_diva_ac_p2,fig:tgv_diva_ac_p3p4}, all configurations tested for ACM-HD led to lower velocity divergence than ACM, which is attributable to the improved convergence properties of hyperbolic diffusion methods. Furthermore, lower values of $T$ could lead to lower values of velocity divergence, but as $\zeta$ increased, this effect diminished. This is indicative of the increased stiffness in the mass equation leading to it becoming dominant in the convergence of the system.
    
    An interesting phenomenon is observed in \cref{fig:diva_en} for EDAC. Here, after peak dissipation, the ratio of absolute divergence to enstrophy becomes approximately linear. Outside of the enstrophy production regime, the source of this can be understood from the vorticity form of the momentum equation. A term that scales with both vorticity and velocity divergence, that would ordinarily cancel, is present in the vorticity equation. The conclusion, which may have been anticipated, is that vortex dominated flows may induce stronger divergence clearing from the method, which may in turn be problematic for EDAC as this manifests as larger pressure fluctuations. At lower order, due to lower enstrophy production and higher numerical dissipation, this relationship is not as clear.
    
    In several places we have considered the maximum eigenvalue due to its importance in the efficacy of the explicit Runge--Kutta time integration. A lower absolute maximum eigenvalue will increase the maximum stable time step, but clearly from \cref{ssec:edac} increasing $\zeta$ will increase the maximum absolute eigenvalue. Therefore, a trade-off between stiffness, runtime and divergence, exists.  
    
    \begin{figure}[tbhp]
        \centering
        \subfloat[Evolution of the ratio of absolute divergence to enstrophy in the TGV. Here a 20 point moving average is applied to $s_a$.]{\label{fig:diva_en}\adjustbox{width=0.47\linewidth, valign=b}{\input{figs/tgv_edac_diva_en}}}
        ~
        \subfloat[Average $\Delta t$ for the the TGV using EDAC, solid: Rusanov, dashed: HLL, dotted: HLLC.]{\label{fig:tgv_dt}\adjustbox{width=0.5\linewidth, valign=b}{\input{figs/tgv_edac_dt}}}
        \caption{}
    \end{figure}
    
    For the EDAC method, a globally adaptive time stepping procedure was used. To understand the trade-off that occurs, the statistics of the time steps used were collected and the average time step size is shown in \cref{fig:tgv_dt}. This shows the same linear relationship at high $\sqrt{\zeta}$ values as was observed for the divergence. In the case of $\Delta t$, the origin of this relationship can be clearly seen form the eigenvalues of the inviscid flux for the EDAC system of equation. In \cref{fig:tgv_edac_diva,fig:tgv_dt} results are also presented for HLL and HLLC as well as Rusanov. The HLL and HLLC schemes are derived in \cref{sec:riemann}. These data show that HLLC can give a sizeable reduction in the divergence average, and both HLL and HLLC lead to an increase in the maximum usable $\Delta t$. The cause of this is a better model of the maximum stable eigenvalue, as the Davis wavespeed estimate more often over predicts this wavespeed~\citep{Toro2020}, which will lead to a stricter CFL condition. Based on these results there is a clear benefit to using the HLLC Riemann solver at the interfaces, and given the availability of FLOPs on GPU, the additional computation can be largely hidden by memory latency caused by bandwidth limitations.

\subsection{Circular Cylinder at $Re=3900$}\label{ssec:cylinder}
    The circular cylinder has previously been studied in great detail, a comprehensive review was presented by \citet{Williamson1996}. When $Re=3900$, the flow is in a sub-critical regime and is considered to be both computationally challenging and physically interesting. This is due to the presence of multiple phenomena, namely: a laminar boundary-layer, separation, a free shear-layer, turbulent transition of the free shear-layer, and a turbulent wake. The difficulties in simulating this case are exemplified by the spread of reported drag coefficients in the literature~\citep{Pereira2018}, with a value in the range of $C_d\in[1,1.4]$ being not uncommon. Subsequently reported by \citet{Lehmkuhl2013}, a cause for this is that the time averaged wake has two distinct modes, a low energy L-mode and a high energy H-mode, with long non-dimensional times separating transition event between the modes, typically of the order $10^3$. Owing to this long time between transitions, a simulation will typically only capture a single mode. 
    
    \begin{figure}[tbhp]
        \centering
        \subfloat[EDAC, $\zeta=4$, $\hat{t}=201$.]{\adjustbox{width=0.3\linewidth, valign=b}{\includegraphics[]{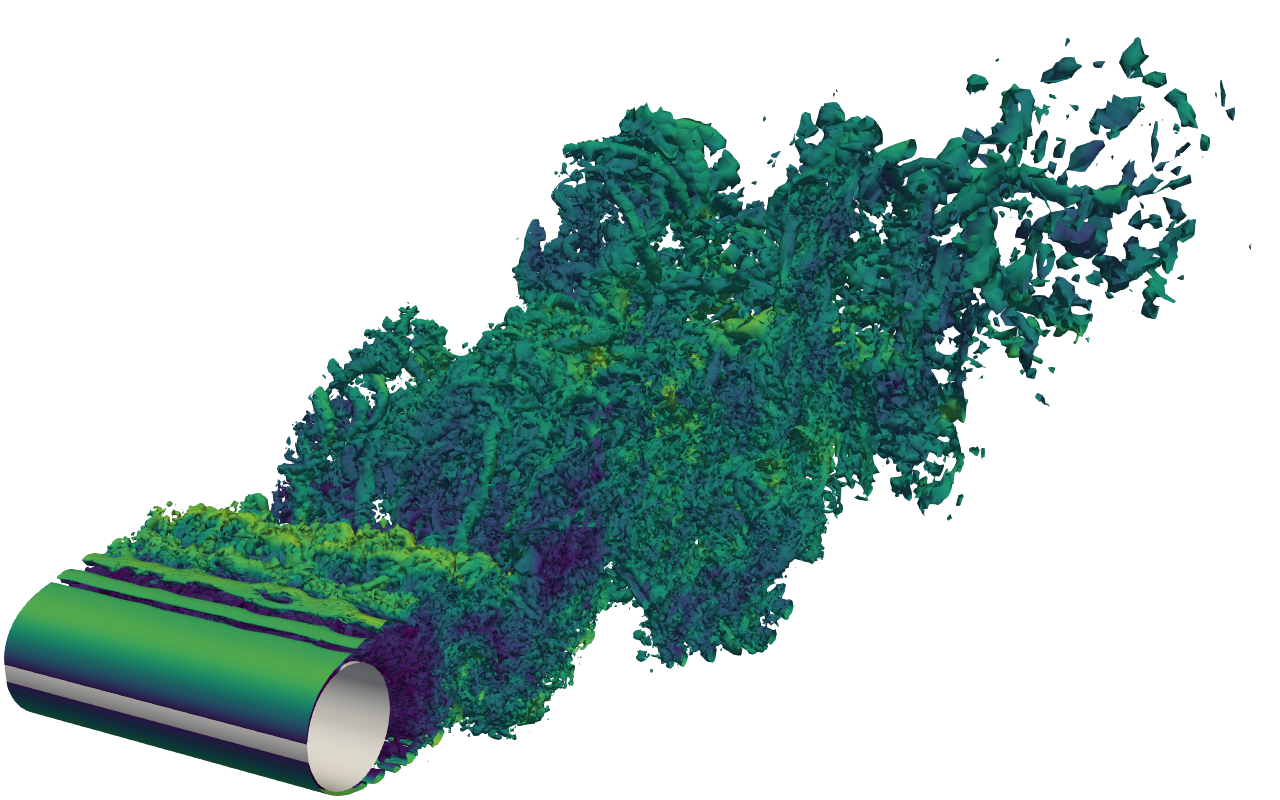}}}
        ~
        \subfloat[EDAC, $\zeta=20$, $\hat{t}=201$.]{\adjustbox{width=0.3\linewidth, valign=b}{\includegraphics[]{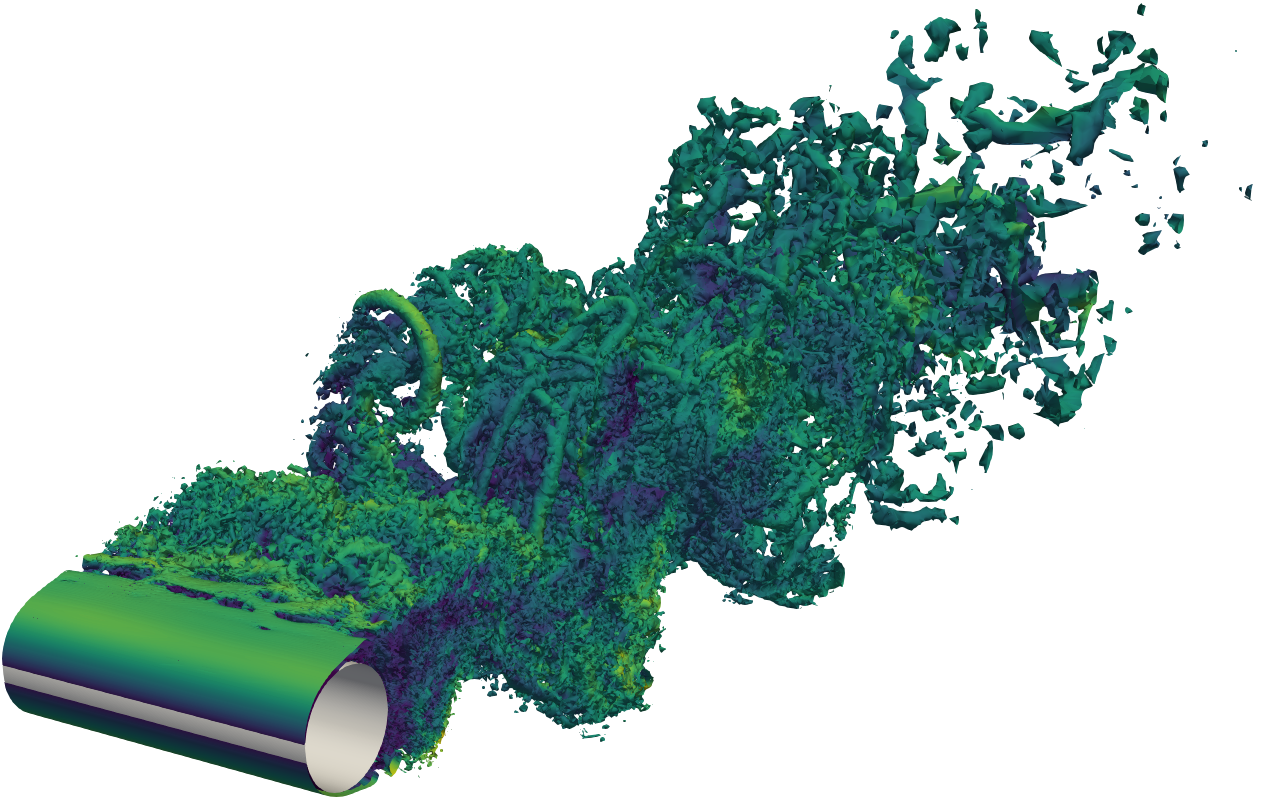}}}
        ~
        \subfloat[EDAC, $\zeta=100$, $\hat{t}=201$.]{\adjustbox{width=0.3\linewidth, valign=b}{\includegraphics[]{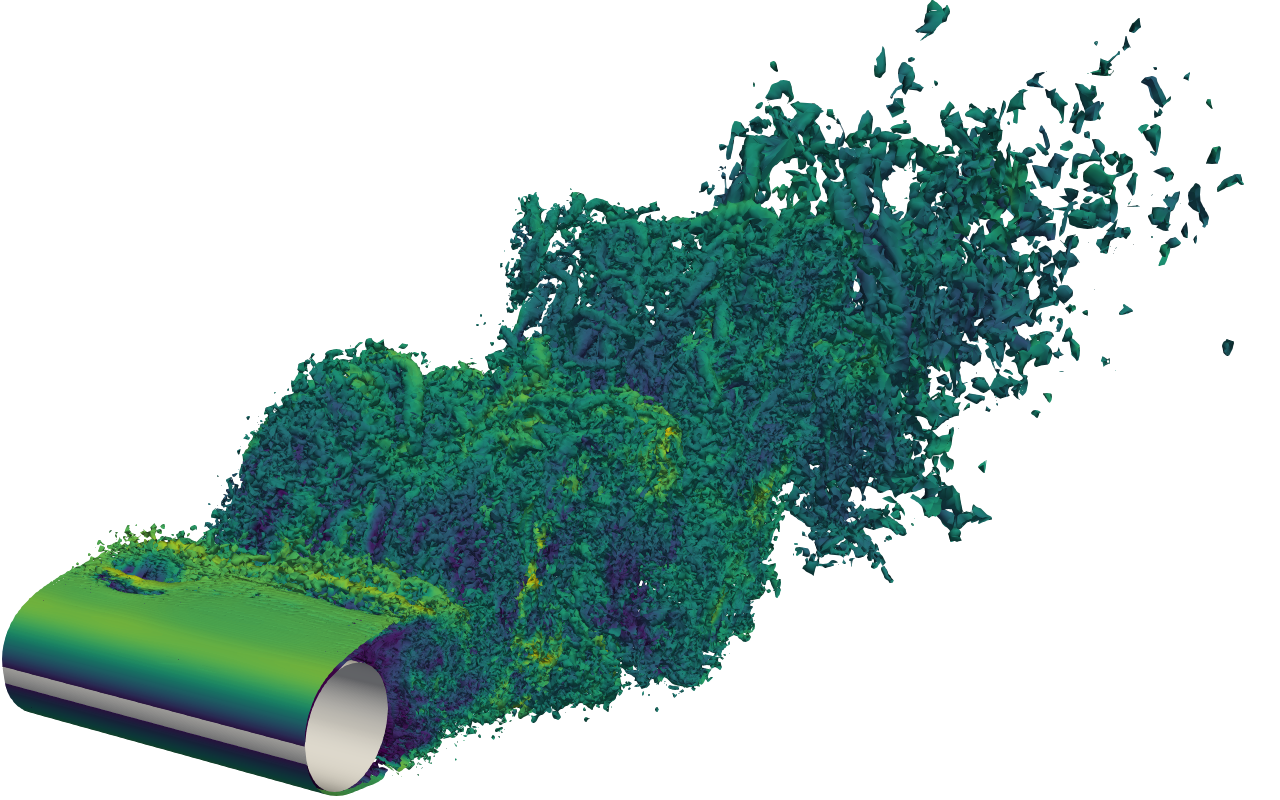}}}
        \caption{\label{fig:cyl-q}Iso-contours of Q-criterion, $Q=0.1$, coloured by velocity magnitude.}
    \end{figure}
    
    In the work of \citet{Vermeire2017} the PyFR solver was compared to other tools. It was shown that with an appropriate mesh simulations of the compressible Navier--Stokes equations at $M=0.2$ gave good agreement with the DNS results of \citet{Lehmkuhl2013}. For this reason we will use the same mesh in this investigation which covered the domain $\Omega = [-9D,25D]\times[-9D,-9D]\times[0,\pi D]$ for diameter $D$. More recently, \citet{Dzanic2021} performed a DNS of this case. Comparison is made later to their data, as well as to additional TKE budget data collected by the present authors rerunning their configuration.

    \begin{figure}[tbhp]
        \centering
        \subfloat[$\avg{u}/U$.]{\label{fig:cyl_u_1p06}\adjustbox{width=0.42\linewidth,valign=b}{\input{figs/cylinder_edac_u_y_x1-06}}}
        ~
        \subfloat[$\avg{v}/U$.]{\label{fig:cyl_v_1p06}\adjustbox{width=0.47\linewidth,valign=b}{\input{figs/cylinder_edac_v_y_x1-06}}}
        \\
        \subfloat[$\avg{u^\prime u^\prime}/U$.]{\label{fig:cyl_upup_1p06}\adjustbox{width=0.42\linewidth,valign=b}{\input{figs/cylinder_edac_upup_y_x1-06}}}
        ~
        \subfloat[$\avg{v^\prime v^\prime}/U^2$.]{\label{fig:cyl_vpvp_1p06}\adjustbox{width=0.42\linewidth,valign=b}{\input{figs/cylinder_edac_vpvp_y_x1-06}}}
        \caption{\label{fig:cylinder_x1-06}$x/D=1.06$, with experimental data from \citet{Parnaudeau2008} and DNS data from \citet{Lehmkuhl2013}.}
    \end{figure}
    
    For this test, seven configurations were tested: EDAC with $\zeta\in\{4,20,100\}$, ACM with $\zeta\in\{3,4.5\}$ and ACM-HD $\zeta=3$ with $T\in\{10\nu, 100\nu\}$. The case was set with a free stream velocity and pressure of $U=0.2\sqrt{\gamma}$ and $P_0=1$. Constant velocity inlet and constant pressure outlet boundary conditions were used for all cases. Although, Riemann invariant boundary conditions are available, they were deemed unnecessary in this case. An initialisation period was set from 0 to $\hat{t}=tU/D=47$ at $p=2$ followed by $p=4$ until $\hat{t}=100$. Afterwards, the time average statistics were collected until $\hat{t}=200$. To accelerate the convergence of the ACM and ACM-HD dual time stepping, the same P-multigrid method was used as described in \cref{ssec:tgv}.
    
    The instantaneous Q-criterion for the EDAC method at $\hat{t}=Ut/D=201$ is shown in \cref{fig:cyl-q}. Some differences in the wake structure are visible and notably at lower $\zeta$ values the free shear layer appears to transition earlier. Focusing on EDAC initially, the time and span averaged velocity profiles --- \cref{fig:cylinder_x1-06,fig:cylinder_x2-02} --- show that there is a pronounced difference between the different $\zeta$ values, with the highest value of $\zeta$ giving the best agreement with the the DNS H-mode. It is interesting that as $\zeta$ is increased monotonic convergence is not observed. From the Q-criterion plots, it seems as though transition is delayed for $\zeta=20$ compared to $\zeta=4$; however, small structures in the wake look as though they have coalesced into larger scale structures. Further comparison can be made with additional Q-criterion plots in \cref{app:cylinder_plots}.
    
    Studying the progression of the streamwise Reynolds stress through the downstream slices, \cref{fig:cyl_upup_1p06,fig:cyl_upup_2p02}, it is observed that EDAC initially overestimates the Reynolds stresses, leading to the downstream stresses at $x/D=2.02$ being under-predicted. Clearly this will have an impact on the turbulent kinetic energy budget, a detailed study of which is presented by \citet{Tian2020}, who showed that at $x/D=1.06$, a balance exists between the convection, production, pressure transport, turbulent transport, and dissipation. Closer still to the cylinder, \citet{Tian2020} showed the production term is dominant, and we hypothesise that over-production would lead to a downstream energy deficit, and that this is happening for EDAC. The turbulent production is defined as
    \begin{equation}
        \mathcal{P} = \avg{u_iu_j}\overline{S}_{ij}, \quad \mathrm{and} \quad \overline{S}_{ij} = \frac{1}{2}\left(\px{\avg{u_i}}{x_j} + \px{\avg{u_j}}{x_i}\right),
    \end{equation}
    where we have used Einstein notation.
    
\begin{figure}[tbhp]
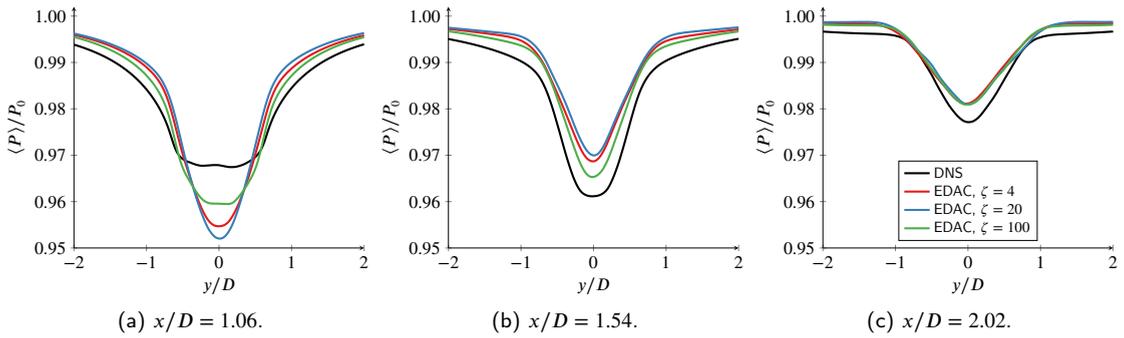

        \centering
        \subfloat[$x/D=1.06$.]{\adjustbox{width=0.3\linewidth, valign=b}{\input{figs/cylinder_pressure_x1.06}}}
        \subfloat[$x/D=1.54$.]{\adjustbox{width=0.3\linewidth, valign=b}{\input{figs/cylinder_pressure_x1.54}}}
        \subfloat[$x/D=2.02$.]{\adjustbox{width=0.3\linewidth, valign=b}{\input{figs/cylinder_pressure_x2.02}}}
        \caption{\label{fig:cyl_pressure}Span averaged pressure, $\avg{P}/P_0$, for several downstream slices, DNS from \citet{Dzanic2021}.}
    \end{figure}

    Plotting the production term at $x/D=1.06$, \cref{fig:cyl_prod_x1-06}, over-production is clearly visible in comparison to DNS data. This is followed by a drop and subsequent under-prediction of production downstream, \cref{fig:cyl_prod_x2-02}. The question arises, what is driving this over production? From the average pressure profiles of \cref{fig:cyl_pressure} it can be seen that the pressure in the wake region is significantly lower than the DNS of \citet{Dzanic2021}. This is a result of the artificial compressibility in the pressure field.  From the interpretation that $\zeta=1/M^2$, it can be understood that this pressure variation will increase as $\zeta$ decreases. The pressure gradient this leads to is the driving force of the increased production. Downstream the pressure gradient reduces and there is an associated drop in production. This highlights the difference between the EDAC method and ACM. In ACM, the velocity divergence is balanced by pressure fluctuations resolved in pseudo-time, whereas the EDAC method uses the alternative mechanism of spatial pressure variations. The spatial pressure fluctuation are able to have a significant impact on the observed physics, whereas in pseudo-time they are diffused and convected out of the solution by the explicit smoother. A further effect of the pressure reduction is an increase in the entropy dissipation, as seen from \cref{eq:enst}, which helps to explain the energy deficit. 
    
    Studying the ACM and ACM-HD results, it is interesting to see that from \cref{fig:cylinder_x1-06}, the ACM simulation with $\zeta=3$ is in the L shedding mode, whereas all the other results are in the H shedding mode, which acts to highlight the sensitivity of this case and which mode is initially captured. The ACM-HD results for $T=100\nu$ outperform those of $T=10\nu$, given that a higher $T$ reduces the stiffness while in the asymptotic limit a higher $T$ is preferred. The ACM results with $\zeta=4.5$ are comparable to ACM-HD with $\zeta=3$ and $T=100\nu$. However, from the case setup, this ACM configuration was on the limit of what is stable. This can also be understood when considering the results of \cref{fig:tgv_diva_ac_p3p4}.
    
    \begin{figure}[tbhp]
        \centering
        \subfloat[$\avg{u}/U$.]{\label{fig:cyl_u_2p02}\adjustbox{width=0.42\linewidth,valign=b}{\input{figs/cylinder_edac_u_y_x2-02}}}
        ~
        \subfloat[$\avg{v}/U$.]{\label{fig:cyl_v_2p02}\adjustbox{width=0.47\linewidth,valign=b}{\input{figs/cylinder_edac_v_y_x2-02}}}
        \\
        \subfloat[$\avg{u^\prime u^\prime}/U$.]{\label{fig:cyl_upup_2p02}\adjustbox{width=0.42\linewidth,valign=b}{\input{figs/cylinder_edac_upup_y_x2-02}}}
        ~
        \subfloat[$\avg{v^\prime v^\prime}/U$.]{\label{fig:cyl_vpvp_2p02}\adjustbox{width=0.42\linewidth,valign=b}{\input{figs/cylinder_edac_vpvp_y_x2-02}}}
        \caption{\label{fig:cylinder_x2-02}$x/D=2.02$, with experimental data from \citet{Parnaudeau2008} and DNS data from \citet{Lehmkuhl2013}.}
    \end{figure}

    \begin{figure}[tbhp]
        \centering
        \subfloat[$x/D=1.06$.]{\label{fig:cyl_prod_x1-06}\adjustbox{width=0.3\linewidth, valign=b}{\input{figs/cylinder_production_x1.06}}}
        \subfloat[$x/D=1.54$.]{\adjustbox{width=0.3\linewidth, valign=b}{\input{figs/cylinder_production_x1.54}}}
        \subfloat[$x/D=2.02$.]{\label{fig:cyl_prod_x2-02}\adjustbox{width=0.3\linewidth, valign=b}{\input{figs/cylinder_production_x2.02}}}
        \caption{\label{fig:cyl_production}Span averaged turbulent production, $\mathcal{P}D/U^3$, for several downstream slices, DNS from \citet{Dzanic2021}.}
    \end{figure}
    
    As was discussed in \cref{ssec:tgv}, the increased stiffness of the EDAC method at higher $\zeta$ values will lead to lower maximum stable time steps which will negatively impact the runtime. The data in \cref{tab:cyl_runtime} gives a runtime comparison for one flow over diameter of the cylinder, from $\hat{t}=201$ to $\hat{t}=202$. This shows that increasing $\zeta$ by a factor of 5 gives approximately doubles the runtime of EDAC, which is approximately in line with the increase in the value of $\lambda$ predicted from section \cref{ssec:edac}. Furthermore, the runtimes for EDAC were observed to be lower than those of ACM; however, to achieve comparable results to ACM, $\zeta$ for EDAC would need to increased and is unlikely to be faster in that case. For the cylinder case with ACM it was found that $\Delta t=\num{2e-3}$ and $\Delta t/\Delta\tau=20$ were stable, whereas due to the increased stability of ACM-HD $\Delta t=\num{4e-3}$ and $\Delta t/\Delta\tau=20$ could be used. This is the cause of reduced runtime of ACM-HD compared to ACM observed in \cref{tab:cyl_runtime}, and is one of the attractive features of hyperbolic diffusion. Although more equations are solved, the simpler advection algorithm coupled to the stability improvement lead to a speedup even at lower Reynolds numbers~\citep{Trojak2021}. 
    
    \begin{table}[tbhp]
        \centering
        \begin{tabular}{l c r}
            \toprule
            Scheme & Configuration & Wall time \\ \midrule
            ACM, $\zeta=4.5$ & RK3(2)4[2R+], PMG, LAPTS & 0:32:14 \\
            ACM, $\zeta=3$ & RK3(2)4[2R+], PMG, LAPTS & 0:31:55 \\
            ACM-HD, $\zeta=3$, $T=100\nu$ & RK3(2)4[2R+], PMG, LAPTS & 0:27:36 \\
            ACM-HD, $\zeta=3$, $T=10\nu$ & RK3(2)4[2R+], PMG, LAPTS & 0:22:29 \\
            EDAC, $\zeta=100$ & RK3(2)4[2R+], GAPTS & 0:25:12 \\
            EDAC, $\zeta=20$ & RK3(2)4[2R+], GAPTS & 0:12:18 \\
            EDAC, $\zeta=4$ & RK3(2)4[2R+], GAPTS & 0:06:48 \\ \bottomrule
        \end{tabular}
        \caption{\label{tab:cyl_runtime}Runtime for circular cylinder at $Re=3900$ and $p=4$ for one flow over diameter and 12 partitions run using NVIDIA V100 GPUs.}
    \end{table}

\subsection{SD7003 at $Re=\num{6e4}$}\label{ssec:sd7003}
    The SD-7003 aerofoil~\citep{Selig1989} is an asymmetric aerofoil that has been explored in several studies, for example the studies by \citet{Beck2014}, \citet{Galbraith2010}, and \citet{Garmann2012}. The particular configuration that will be tested here is that of $Re=\num{6e4}$ and an angle of attack of $\alpha=8^\circ$. In this configuration, a laminar separation bubble is formed on the suction surface followed by turbulent transition of the boundary layer. From numerical studies~\citep{Beck2014,Galbraith2010,Garmann2012}, it has been shown that the lift and drag coefficient are sensitive to the separation location, which in turn will be dependent on the surface pressure distribution. This configuration was previously used by \citet{Loppi2019} to assess the effectiveness of locally adaptive pseudo-time stepping and P-multigrid in the convergence acceleration of ACM within the PyFR framework. 
    
    In order to allow direct comparison to there results, the same mesh was used in this investigation. As it was found in \citet{Loppi2019} to give good agreement with DNS data. The mesh is made up of \num{137916} hexahedral elements covering a domain formed of a rectangular downwind section of $[0,20c]\times[-10c,10c]\times[0,0.2c]$, with an upwind extruded semi-circle of diameter $20c$, where $c$ is the chord length. The time integration method used here is again an explicit RK3(2)4[2R+] scheme with globally adaptive time stepping. The calculation was performed for $t\in[0,60]$ with the divergence and global force metrics output periodically as well as data on the time step size used. As this case is at a higher $Re$, aliasing is of greater concern. To ameliorate this, over-integration anti-aliasing of the flux at the solution points is used~\citep{Spiegel2015}. All methods on this case were run at $p=4$ and the over-integration order was set to $q=11$, similar to the setup of \citet{Loppi2019}.
    
    \begin{figure}[tbhp]
        \centering
        \subfloat[EDAC, $\zeta=4$, $\hat{t}=47$.]{\adjustbox{width=0.3\linewidth, valign=b}{\includegraphics[]{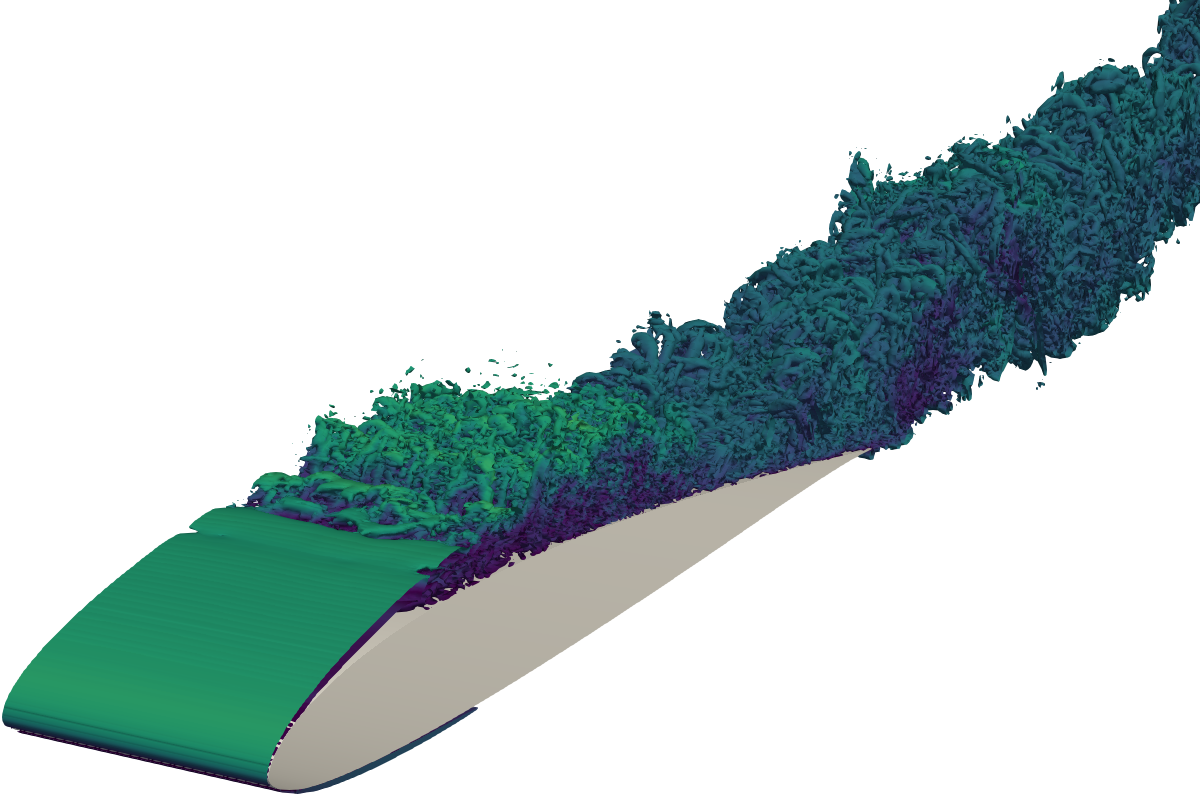}
        }}
        ~
        \subfloat[EDAC, $\zeta=20$, $\hat{t}=47$.]{\adjustbox{width=0.3\linewidth, valign=b}{\includegraphics[]{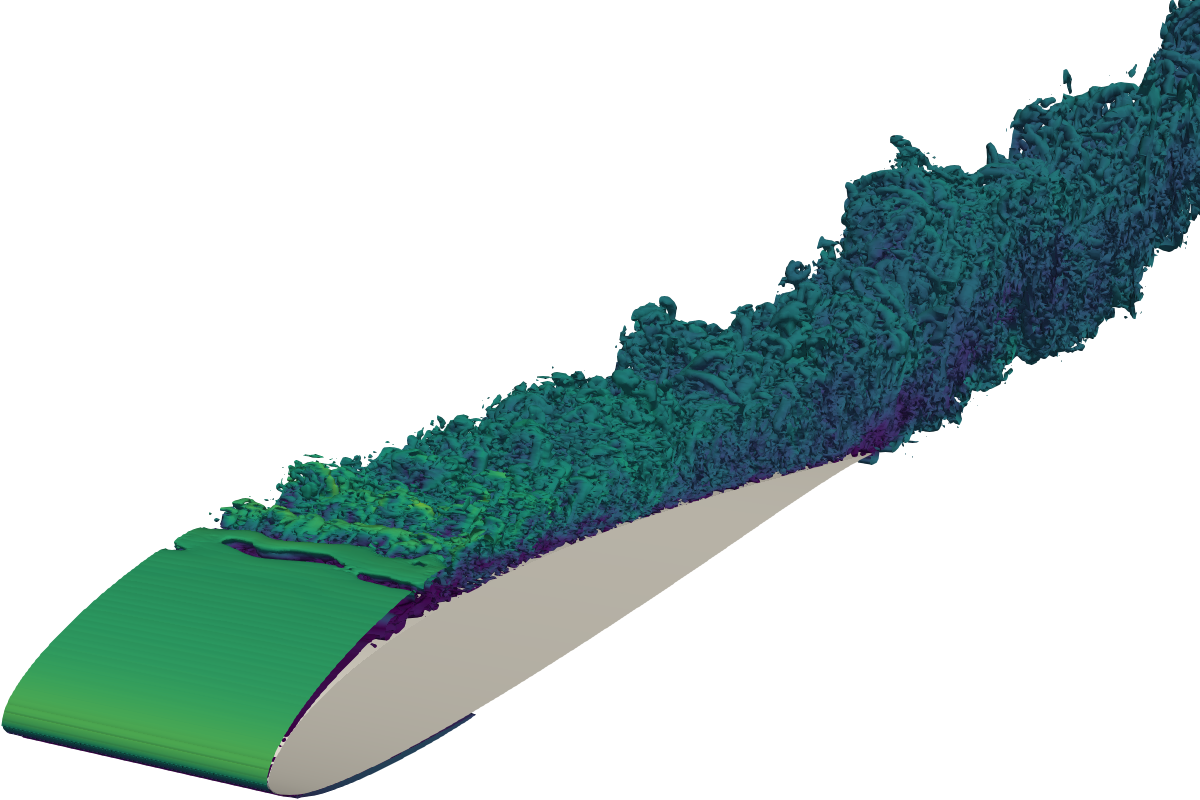}}}
        ~
        \subfloat[EDAC, $\zeta=100$, $\hat{t}=47$.]{\adjustbox{width=0.3\linewidth, valign=b}{\includegraphics[]{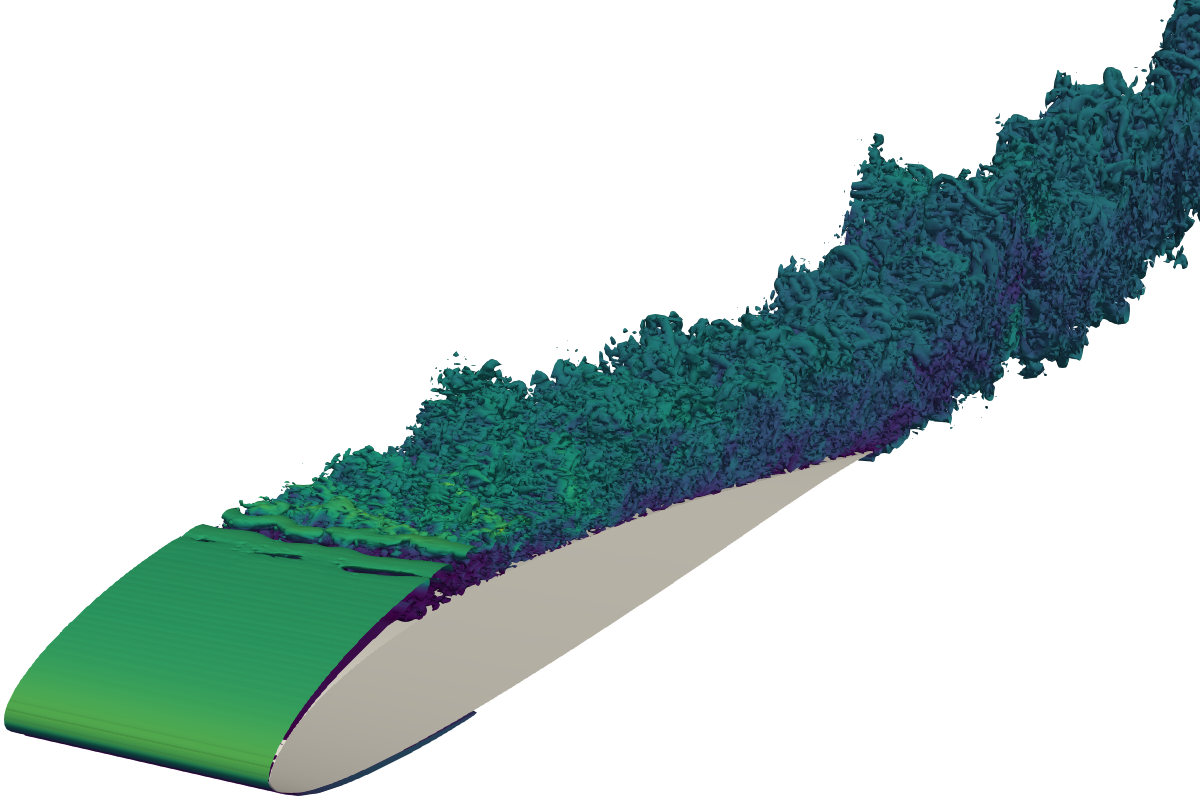}}}
        \caption{\label{fig:sd-q}Iso-contours of Q-criterion, $Q=20$, coloured by velocity magnitude for the SD7003 at $Re=\num{6e4}$ and $\alpha=8^\circ$.}
    \end{figure}

\begin{figure}[tbhp]
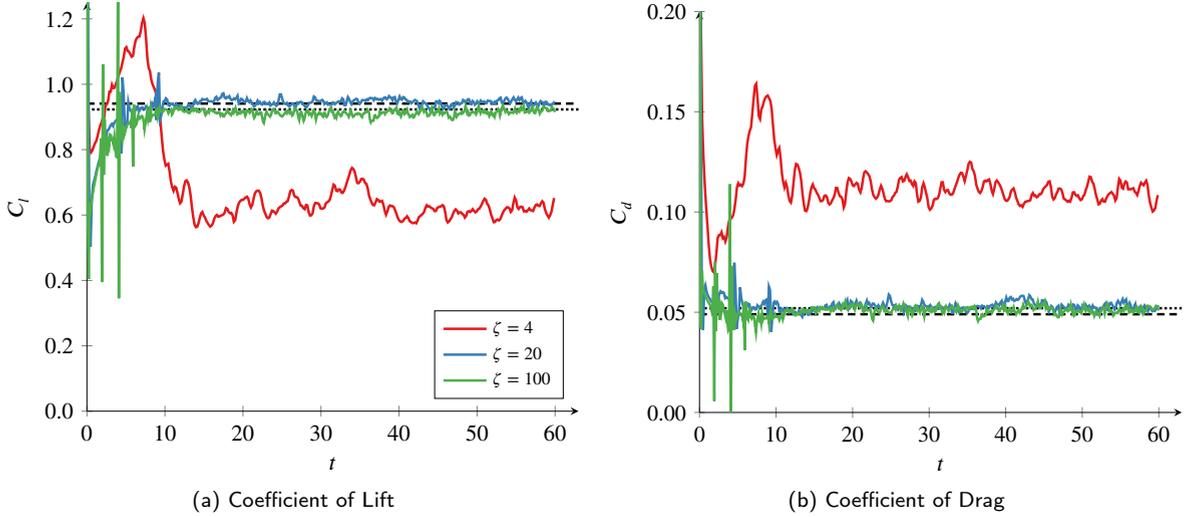

        \centering
        \subfloat[Coefficient of Lift]{\label{fig:sd7003_cl}\adjustbox{width=0.47\linewidth,valign=b}{\input{figs/sd7003_cl_comp}}}
        ~
        \subfloat[Coefficient of Drag]{\label{fig:sd7003_cd}\adjustbox{width=0.47\linewidth,valign=b}{\input{figs/sd7003_cd_comp}}}
        \caption{\label{fig:sd7003_clcd}Coefficient of lift and drag, using FR, $p=4$, for the SD7003 at $\alpha=8^\circ$ and $Re=\num{6e4}$. (dotted) data for ACM on the same mesh by \citet{Loppi2019}, (dashed) is data at $M=0.2$ from \citet{Vermeire2017}.}
    \end{figure}

    The effect of different $\zeta$ terms for the EDAC scheme can be qualitatively understood from the Q-criterion iso-surfaces shown in \cref{fig:sd-q}. From these plots, it looks as though the boundary layer is significantly thicker in the $\zeta=4$ case. There also seems to be some variation in the transition point. Then moving on to study the coefficient of lift and drag, presented in \cref{fig:sd7003_clcd}, it is observed that $\zeta=4$ significantly over-predicts drag and under-predicts lift, consistent with the boundary layer thickening observed in the instantaneous flow fields. At the high values of $\zeta$, good agreement is achieved in comparison to the numerical results of \citet{Vermeire2017} and \citet{Loppi2019}. Subjectively, the lower $\zeta=20$ results seems to be more similar to the compressible results of \citet{Vermeire2017}, whereas $\zeta=100$ compares more favourably with the ACM 
    results of \citet{Loppi2019}.
    
\begin{figure}[tbhp]
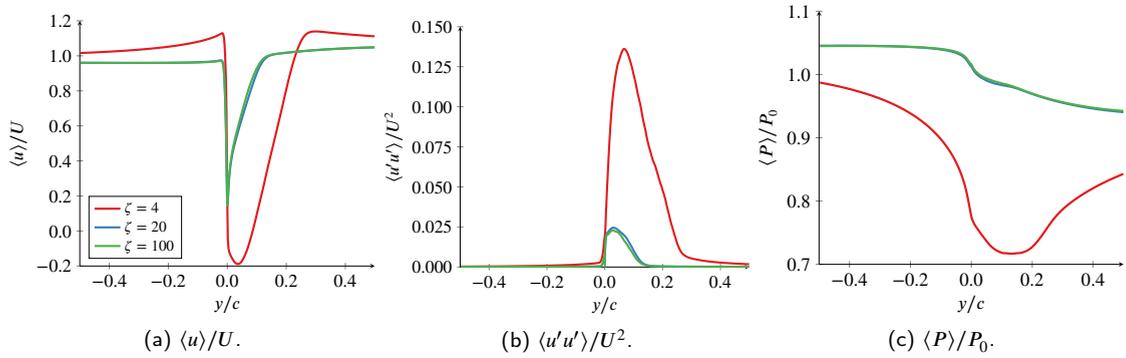

        \centering
        \subfloat[$\avg{u}/U$.]{\label{fig:sd_u}\adjustbox{width=0.3\linewidth, valign=b}{\input{figs/sd7003_slice_u}}}
        \subfloat[$\avg{u^\prime u^\prime}/U^2$.]{\adjustbox{width=0.3\linewidth, valign=b}{\input{figs/sd7003_slice_upup}}}
        \subfloat[$\avg{P}/P_0$.]{\adjustbox{width=0.3\linewidth, valign=b}{\input{figs/sd7003_slice_p}}}
        \caption{\label{fig:sd7003_mean}Span and time averaged qualities at trailing edge of SD7003 at $Re=\num{6e4}$ and $\alpha=8^\circ$. The trailing edge is centred at $y/c =0$.}
    \end{figure}

    To gain greater insight into the cause of the poor results at $\zeta=4$ we studied the time and span averaged stream-wise velocity and pressure at the trailing edge, shown in \cref{fig:sd7003_mean}, averaged over $\hat{t}\in [0,60]$. We were unable to find any detailed DNS data for this particular configuration and location; however, it is clear that there are significant differences in the physics observed. As was observed in \cref{ssec:cylinder}, the spatial pressure variations of EDAC are providing the mechanism driving the inaccuracy. Here, rather than early transition and over-production, \cref{fig:sd_u} shows that there is recirculating region on the suction surface, indicative of a larger adverse pressure gradient. This recirculating flow is then responsible for a thicker boundary layer, which in turn will lead to higher $C_d$. As the thicker boundary layer will reduce the flow turning, this will also lower the $C_l$. 
    
    As was stated earlier, $\zeta$ can have a significant effect on the stiffness of EDAC and $\Delta t$ and the runtime. 
    The average $\Delta t$ for the different EDAC configurations are presented in \cref{tab:sd7003_dt}. Also included in this table is the ratio predicted from using the approximate absolute eigenvalue based on the inflow and outflow. This approximation under-predicts the impact of $\zeta$; however, it does give a good first approximation if trying to predict simulation cost and is consistent with \cref{fig:tgv_dt}.
    
    Similarly to the cylinder test case, this case was run from $\hat{t}=45$ to $\hat{t}=46$, and the runtime recorded, similar to the test performed by \citet{Loppi2019}; however, here NVIDIA V100 GPUs are used. The results are presented in \cref{tab:sd7003_rt} and there is a clear benefit to EDAC over ACM in this case, with ACM-HD being significantly slower than ACM. The reason for this is, although a larger time step can be taken, the anti-aliasing used greatly increases the cost. This is due to the large number of global reads and writes required for the flux calculation. In this case at $p=4$, $q=11$ anti-aliasing is used, given the flux for ACM-HD on a hexahedral element requires $d(1 + d + d^2)(q+1)^d$ values per element, flux anti-aliasing with a high degree will require substantially more bandwidth than ACM. 
    
    \begin{figure}[tbhp]
        \centering
        \adjustbox{width=0.47\linewidth,valign=b}{\input{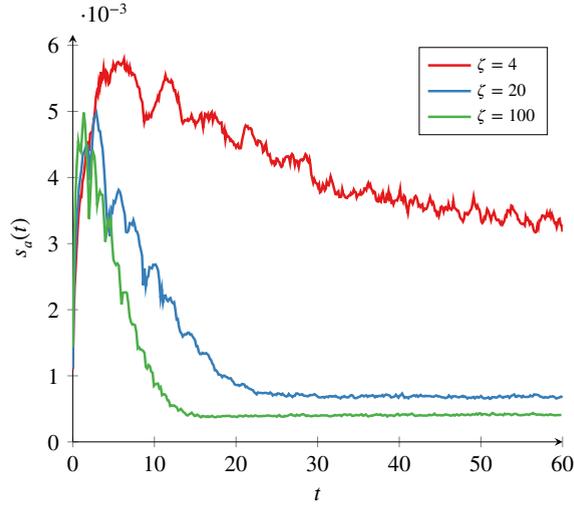}}
        \caption{\label{fig:sd7003_div}SD7003 at $\alpha=8^\circ$ and $Re=\num{6e4}$, volume averaged absolute divergence with time.}
    \end{figure}
    
    \begin{table}[tbhp]
        \centering
        \begin{tabular}{l r r r}
            \toprule
            $\zeta$ & $\overline{\Delta t}$ & $\overline{\Delta t}$ growth & Est. $\lambda_\mathrm{max}$ ratio\\ \midrule
            100 & \num{9.44e-6} & - & -\\
            20 & \num{1.88e-5} & 1.99 & 1.79\\
            4 & \num{3.36e-5} & 1.79 & 1.61\\ \bottomrule
        \end{tabular}
        \caption{\label{tab:sd7003_dt}EDAC average time step size with RK34 for SD7003 case.}
    \end{table}
    
    \begin{table}[tbhp]
        \centering
        \begin{tabular}{l c r}
            \toprule
            Scheme & Configuration & Wall time\\ \midrule
            ACM, $\zeta=3$ & RK3(2)4[2R+], PMG, LAPTS & 1:27:01 \\
            ACM-HD, $\zeta=3$, $T=100\nu$ & RK3(2)4[2R+], PMG, LAPTS & 4:06:31 \\
            EDAC, $\zeta=100$ & RK3(2)4[2R+], GAPTS & 0:45:28 \\
            EDAC, $\zeta=20$ & RK3(2)4[2R+], GAPTS & 0:22:46 \\
            EDAC, $\zeta=4$ & RK3(2)4[2R+], GAPTS & 0:12:46 \\
            \bottomrule
        \end{tabular}
        \caption{\label{tab:sd7003_rt}SD7003 run-time comparison for partition of 32 NVIDIA V100 GPUs.}
    \end{table}

\section{Conclusions}\label{sec:conclusions}
    In this work we explored the properties of EDAC, the widely used ACM technique, and the hyperbolic-diffusion approach ACM-HD. Novel insight into the Riemann problem for the conservative EDAC system and ACM-HD system was provided and displayed the effect the various parameters of the schemes would have on stiffness. This analysis identified some unique challenges when formulating either maximum absolute eigenvalues estimations or Riemann invariant boundary conditions in EDAC and ACM-HD. 
    
    The main advantage of the EDAC system is the ability to use explicit time stepping when simulating the unsteady incompressible Navier--Stokes equations. It was shown in simulations of the Taylor--Green vortex that as the EDAC compressibility parameter, $\zeta$ tended to infinity, the divergence scaled linearly with $\sqrt{\zeta}$. Furthermore, it was found that HLLC common interface calculations could significantly reduce divergence and increase the maximum stable time step compared to Rusanov. In contrast, for ACM and ACM-HD as $\zeta$ tended to infinity this linear relation was not observed, instead improvements in divergence diminished. It was shown that in more challenging LES cases EDAC can lead to erroneous physics if incorrectly configured. For the circular cylinder at $Re=\num{3900}$, the free shear layer transition could be triggered early by the spatial pressure variation that balances divergence in EDAC. Errors in transitions were also observed with EDAC if poorly configured on the SD7003 aerofoil at  $Re=\num{6e4}$. Again if $\zeta$ was too small, the spatial pressure variations could prevent boundary layer reattachment. ACM and ACM-HD simulations on the other hand did not face this problem, with ACM-HD having a comparable or faster run time than EDAC on the cylinder case. The excessive stiffness of the EDAC system required to correctly simulate transition was found to be responsible for long runtime.
    
    It is concluded that EDAC can be an effective alternative to ACM for unsteady problems, although care must be taken in transitional cases. Furthermore, the improved stability and convergence properties of ACM-HD can make more favourable compared to ACM.
    
\section*{Acknowledgements}\label{sec:ack}
    This material is based upon work supported by, or in part by, the U. S. Air Force Office of Scientific Research DURIP program under FA9550-21-1-0190 for Enabling next-generation heterogeneous computing for massively parallel high-order compressible CFD, under the direction of Fariba Fahroo. The authors would like to gratefully acknowledge Tarik Dzanic for providing DNS data of the cylinder and proofreading, as well as Geng Tian and Zuoli Xiao for making their data available, although we did not use this in the final manuscript. 

\bibliographystyle{cas-model2-names}
\bibliography{reference}


\clearpage
\begin{appendices}
\section{ACM-HD Riemann Problem}\label{sec:riemann_acmhd}
\begin{figure}[tbhp]
        \centering
        \adjustbox{width=0.5\linewidth,valign=b}{\input{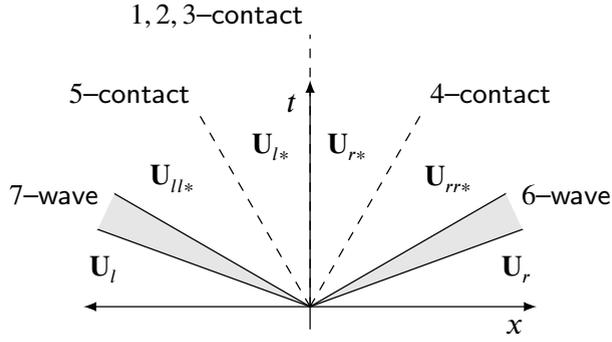}}
        \caption{\label{fig:riemann_fan_acmhd}Riemann fan for 2D ACM-HD.}
    \end{figure}

    To understand the structure and limitations of the Riemann problem for the ACM-HD system we focus on the 2D system. Calculating the inviscid flux Jacobian in the $x$ direction as:
    \begin{equation}
        \px{\mathbf{f}}{\mathbf{U}} = \begin{bmatrix}
            0 & \zeta & 0 & 0 & 0 & 0 & 0 \\
            1 & 2u & 0 & -\nu & 0 & 0 & 0 \\
            0 & v & u & 0 & 0 & -\nu & 0 \\
            0 & -1/T & 0 & 0 & 0 & 0 & 0 \\
            0 & 0 & 0 & 0 & 0 & 0 & 0 \\
            0 & 0 & -1/T & 0 & 0 & 0 & 0 \\
            0 & 0 & 0 & 0 & 0 & 0 & 0
        \end{bmatrix}.
    \end{equation}
    This has the eigenvalues:
    \begin{equation}\label{eq:acmhd_lambda}
        \lambda_1 = \lambda_2 = \lambda_3 = 0, \quad \lambda_{4}=\half u + c, \quad \lambda_{5} = \half u - c, \quad \lambda_{6} = u + b, \quad \mathrm{and} \quad \lambda_7 = u - b,
    \end{equation}
    for $c^2=u^2/4 + \nu/T$ and $b^2 = u^2 + \zeta + \nu/T$. The associated eigenvectors are
    \begin{equation*}
        \mathbf{R}^{(1)} = 
            \begin{bmatrix}
                0 \\ 0 \\ 0 \\ 0 \\ 0 \\ 0 \\ 1
            \end{bmatrix}, \quad 
        \mathbf{R}^{(2)} = 
            \begin{bmatrix}
                0 \\ 0 \\ 0 \\ 0 \\ 1 \\ 0 \\ 0
            \end{bmatrix}, \quad 
        \mathbf{R}^{(3)} = 
            \begin{bmatrix}
                \nu \\ 0 \\ 0 \\ 1 \\ 0 \\ 0 \\ 0
            \end{bmatrix}, \quad
        \mathbf{R}^{(4)} = 
            \half\begin{bmatrix}
                0 \\ 0 \\ -T(u+2c) \\ 0 \\ 0 \\ 2 \\ 0
            \end{bmatrix}, \quad
        \mathbf{R}^{(5)} = 
            \half\begin{bmatrix}
                0 \\ 0 \\ -T(u-2c) \\ 0 \\ 0 \\ 2 \\ 0
            \end{bmatrix}.
    \end{equation*}
    \begin{equation*}
        \mathbf{R}^{(6)} = 
            \begin{bmatrix}
                \zeta \\ u+b \\ \tilde{v}(u+b)^2/(a^2+bu) \\ -1/T \\ 0 \\ -v(u+b)/T(a^2+bu) \\ 0
            \end{bmatrix}, \quad
        \mathbf{R}^{(7)} = 
            \begin{bmatrix}
                \zeta \\ u-b \\ v(u-b)^2/(a^2-bu) \\ -1/T \\ 0 \\ -v(u-b)/T(a^2-bu) \\ 0
            \end{bmatrix}, \quad,
    \end{equation*}
    where $a = \sqrt{u^2 + \zeta}$. This shows that the linear degeneracy indicated by the repeated eigenvalues $\lambda_{1}=\lambda_{2}=\lambda_{3}$, results in a stationary contact discontinuity. The discontinuous properties are $q_y$, $r_y$, as well as $P$ and $q_x$. These final two are part of the same wave and so are related through a Riemann invariant. A stationary contact discontinuity makes formulating an exact Riemann solver more complex, although possible. If the purpose of the exact Riemann solver is to form an upper bound on the maximum absolute eigenvalue~\citep{Guermond2016}, then this is not an issue; however, it is not possible to produce a common interface value. A similar issue would be confronted by the HLLC method and a linearisation of the central contact would be necessary. The structure of the 2D ACM-HD Riemann problem is shown in \cref{fig:riemann_fan_acmhd}, given that:
    \begin{equation}
            \lambda_7 \leqslant \lambda_5 \leqslant 0 \leqslant \lambda_4 \leqslant \lambda_6.
    \end{equation}
    
\section{EDAC Riemann Problem}\label{sec:riemann}
\begin{figure}[tbhp]
        \centering
        \adjustbox{width=0.5\linewidth,valign=b}{\input{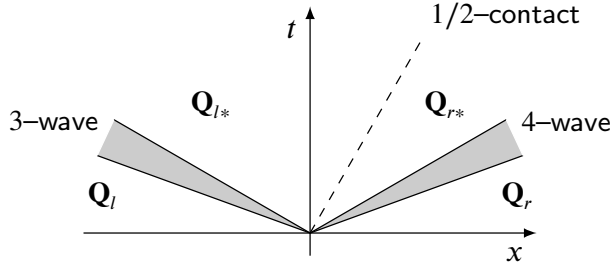}}
        \caption{\label{fig:riemann_fan}Riemann fan for conservative EDAC formulation.}
    \end{figure}

    \noindent
    To understand the structure of the wave fan for EDAC and devise Riemann solvers, we will first state the Riemann problem:
    \begin{equation}
        \pxvar{\mathbf{Q}}{t} + \pxvar{(\mathbf{F}^\mathbf{inv}\cdot\mathbf{n})}{x} = 0, \quad \mathrm{for} \quad \mathbf{n}\in\mathcal{B}(1,0), \quad \mathrm{and}\quad \mathbf{Q}_0 = \begin{cases}
            \mathbf{Q}_L, &\mbox{if } x<0,\\
            \mathbf{Q}_R, &\mbox{otherwise},
        \end{cases}
    \end{equation}
    Using a transformation to the reference problem $\hat{\mathbf{n}}=[1,0,0]^T$, we can form general solutions to the Riemann problem by only studying the first column vector for the invsicid flux, $\mathbf{f}=\mathbf{F}^\mathrm{inv}\cdot\hat{\mathbf{n}}$. The Jacobian of the conservative formulation of EDAC can be straightforwardly found as
    \begin{equation}
        \px{\mathbf{f}}{\mathbf{Q}} = \begin{bmatrix}
            u & P+ \zeta & 0 & 0\\
            1 & 2u & 0 & 0\\
            0 & v & u & 0\\
            0 & w & 0 & u
        \end{bmatrix},
    \end{equation}
    which has the eigenvalues
    \begin{equation}
        \lambda_1=\lambda_2 = u, \quad \lambda_{3} = \frac{3}{2}u - d, \quad \mathrm{and} \quad \lambda_{4} = \frac{3}{2}u + d,
    \end{equation}
    for $d^2 = u^2/4 + P + \zeta$. The associated right eigenvectors are:
    \begin{equation}
        \mathbf{K}_1 = \begin{bmatrix}
            0 \\ 0 \\ 0 \\ 1
        \end{bmatrix}, \quad
        \mathbf{K}_2 = \begin{bmatrix}
            0 \\ 0 \\ 1 \\ 0
        \end{bmatrix}, \quad
        \mathbf{K}_3 = \begin{bmatrix}
            P + \zeta \\ u/2 - d\\ v \\ w
        \end{bmatrix}, \quad
        \mathbf{K}_4 = \begin{bmatrix}
            P + \zeta \\ u/2 + d\\ v \\ w
        \end{bmatrix}.
    \end{equation}
    This combination of eigenvalues and eigenvectors tells use that the Riemann problem for this case does have a linear degeneracy, in that it can support a contact discontinuity in $v$ and $w$, the structure can be seen more clearly in \cref{fig:riemann_fan}. With this established, the strategy to form an exact Riemann solver will follow a similar procedure to \citet{Elsworth1992}. We will include the main steps of the procedure as there are some interesting differences compared to the standard ACM formulation. 
    
    \subsection{Exact Solution}
    The first stage of the exact Riemann solver is to find the non-linear equation that governs $u_*$, this has the general form  
    \begin{equation}\label{eq:riemann_balance}
        0 = P_L - P_R + f_L(u_*) + f_R(u_*) = F_*(u_*)
    \end{equation}
    where $f_L(u_*) = P_* - P_L$, and $f_R(u_*) = P_R - P_*$. Solving this will give $u_*$ and $P_*$. The $i$-waves $3$ and $4$ can be either rarefactions or shock waves and $f$ will depend on the wave type. We will start by finding $f$ for the case when the waves are rarefactions.
    
    \subsubsection{Rarefaction}
    Taking the Riemann invariant across the 3-wave, we find that:
    \begin{equation}
        \dx{P}{u} = -\left(d+\frac{u}{2}\right),
    \end{equation}
    This has the general solution:
    \begin{equation}
        P = C^2 + Cu - \zeta,
    \end{equation}
    for some constant $C$. There is a special case when $C=-u/2$ and hence we find that:
    \begin{equation}
            P_* - P_L = \frac{1}{4}(u_L^2 - u_*^2), \quad \mathrm{and} \quad \dx{}{u_*}(P_* - P_L) = -\frac{1}{2}u_*.
    \end{equation}
    This process can be repeated for the 4-wave, to give a similar solution
    \begin{equation}
        \dx{P}{u} =  d - u/2 \quad \implies \quad P = -\frac{u^2}{4} - \zeta,
    \end{equation}
    which yields
    \begin{equation}
        P_R - P_* = \frac{1}{4}(u_*^2 - u_R^2), \quad \mathrm{and} \quad \dx{}{u_*}(P_R - P_*) = -\frac{1}{2}u_*.
    \end{equation}
    
    \subsubsection{Shock}
    For the shock case, we apply the Rankine--Hugoniot condition, therefore considering the 3-wave we have
    \begin{subequations}
        \begin{align}
            S_L &= \frac{u_L(P_L+\zeta) - u_*(P_* + \zeta)}{P_L - P_*}\\
            S_L &= \frac{u_L^2 + P_L - u_*^2 - P_*}{u_L - u_*}
        \end{align}
    \end{subequations}
    Eliminating $S_L$ and factoring in terms of $P_*-P_L$ we obtain the quadratic
    \begin{equation}
        (P_* - P_L)^2 + u_L(u_* - u_l)(P_* - P_L) - (P_L+\zeta)(u_* - u_L)^2 = 0.
    \end{equation}
    Solving this gives solutions:
    \begin{equation*}
        P_* - P_L = (u_*-u_l)\left[-\frac{u_L}{2}\pm d_L \right]
    \end{equation*}
    To select the physical solution, we use consider the entropy condition across the 3-wave, implying that
    \begin{equation*}
        \frac{3}{2}u_L - d_L > S_L > \frac{3}{2}u_* - d_*.
    \end{equation*}
    This is only satisfied by one of the solutions, therefore we find that
    \begin{equation}
        P_* - P_L = -(u_*-u_l)\left(\frac{u_L}{2} + d_L\right), \quad \mathrm{and}
        \dx{}{u_*}(P_* - P_L) = -\left(\frac{u_L}{2} + d_L\right).
    \end{equation}
    Repeating this process for the 4-wave with the alternative entropy condition of 
    \begin{equation*}
        \frac{3}{2}u_R + d_R > S_R > \frac{3}{2}u_* + d_*,
    \end{equation*}
    we find that
    \begin{equation}
        P_R - P_* = (u_R-u_*)\left(-\frac{u_R}{2} + d_R \right), \quad \mathrm{and} \quad
        \dx{}{u_*}(P_R - P_*) = \left(\frac{u_R}{2} - d_R\right).
    \end{equation}
    
    The condition on whether a wave is a shock or rarefaction is based on the wavespeeds either side of the i-wave. This gives
    \begin{equation}
        \mathrm{Left\; wave} = \begin{cases}
            \mathrm{Rarefaction} &\mbox{if } \lambda_3(u_L, P_L) < \lambda_3(u_*, P_{*L}),\\
            \mathrm{Shock} &\mbox{otherwise},
        \end{cases}
    \end{equation}
    and
    \begin{equation}
        \mathrm{Right\; wave} = \begin{cases}
            \mathrm{Rarefaction} &\mbox{if } \lambda_4(u_*, P_{*R}) < \lambda_4(u_R, P_{*R}),\\
            \mathrm{Shock} &\mbox{otherwise}.
        \end{cases}
    \end{equation}
    To solve \cref{eq:riemann_balance} Newton's method can be applied and in the limit of convergence $P_{*L}=P_{*R}=P_*$. 
    
    \subsection{HLL and HLLC Riemann Solver}
        A common approach taken to design a low dissipation approximate Riemann solver is to use the HLL and HLLC approaches first described by \citet{Harten1983}. The HLL approach  assumes that all variables are constant in the region that is bound by the extreme i-waves. The speeds of these waves is approximated using a Davis approximation as
        \begin{equation}
            S_L = \min{\left(\frac{3}{2}u_L - d_L, \frac{3}{2}u_R - d_R\right)}, \quad \mathrm{and}\quad
            S_R = \max{\left(\frac{3}{2}u_L + d_L, \frac{3}{2}u_R + d_r\right)},
        \end{equation}
        and the star flux takes the standard form:
        \begin{equation}
            \mathbf{f}_* = \frac{S_R\mathbf{f}_L  - S_L\mathbf{f}_R + S_LS_R(\mathbf{u}_R - \mathbf{U}_L)}{S_R - S_L}.
        \end{equation}
        This gives the HLL reconstruction as
        \begin{equation}
            \mathbf{f}_\mathrm{HLL} = \begin{cases}
                \mathbf{f}_L &\mbox{if } 0\leqslant S_L\\
                \mathbf{f}_* &\mbox{if } S_L<0<S_R\\
                \mathbf{f}_R &\mbox{if } S_R\leqslant
            \end{cases}
        \end{equation}
        
        The HLLC approach instead models the contact discontinuity, however in the star region ${u}$ and ${P}$ are constant, hence, we will use the HLL method to approximate $u_*$ and $P_*$. Therefore taking:
        \begin{equation}
            P_{*L} = P_{*R} = P_*, \quad \mathrm{and} \quad u_{*L} = u_{*R} = S_*,
        \end{equation}
        then From the Rankine--Hugoniot condition we have that
        \begin{subequations}
            \begin{align}
                \mathbf{f}_{*L} &= S_L(\mathbf{U}_{*L} - \mathbf{U}_L) + \mathbf{f}_L, \\
                \mathbf{f}_{*R} &= S_R(\mathbf{U}_{*R} - \mathbf{U}_R) + \mathbf{f}_R.
            \end{align}
        \end{subequations}
        These can be used to find the closed relations for $S_*$ and $P_*$ as
        \begin{equation}
            S_* = \frac{S_LS_R(P_L-P_R) - S_Ru_L(P_L + \zeta) + S_Lu_R(P_R + \zeta)}{(S_L - u_L)(P_L + \zeta) - (S_R - u_R)(P_R + \zeta)}, \quad \mathrm{and} \quad P_* = \left(\frac{S_L - u_L}{S_L - S_*}\right)\left(P_L + \left[\frac{S_* - u_L}{S_L - u_L} \right]\zeta\right).
        \end{equation}
        The tangential velocity components can then be found as 
        \begin{equation}
            v_{*x} = \left(\frac{S_x - u_x}{S_x - S_*}\right)v_x, \quad \mathrm{and} \quad w_{*x} = \left(\frac{S_x - u_x}{S_x - S_*}\right)w_x,
        \end{equation}
        where $x$ is $L$ or $R$.
        The complete star states can then be written as
        \begin{equation}
            \mathbf{U}_{*x} = \Gamma_x\begin{bmatrix}
                P_L + \zeta(S_* - u_x)/(S_x - u_x) \\ S_*/\Gamma_x \\ v_x \\ w_x
            \end{bmatrix}, \quad \mathrm{for}\quad \Gamma_x = \frac{S_x - u_x}{S_x - S_*},
        \end{equation}
        and the final scheme is
        \begin{equation}
            \mathbf{f}_\mathrm{HLLC} = \begin{cases}
                \mathbf{f}_L &\mbox{if } 0\leqslant S_L \\
                \mathbf{f}_{*L} &\mbox{if } S_L < 0 \leqslant S_* \\
                \mathbf{f}_{*R} &\mbox{if } S_* < 0 < S_R \\
                \mathbf{f}_R &\mbox{if } S_R \leqslant 0
            \end{cases}
        \end{equation}
        
        The most straightforward means to apply these approximate Riemann solvers is by transforming the solution from the normal $\hat{\mathbf{n}}$ to a reference normal, $[1,0,0]^T$, calculating $\mathbf{f}$ and then transforming back again. In two-dimensions this is trivial, however, in three-dimensions this transformation can suffer from numerical errors. A fix for this issue is the approach of \citet{Moller1999}, there Householder transformations are used when the normal vector is closely aligned to either of the tangential cardinal directions.
    
    \subsection{Riemann Invariant Boundary Condition}
        Now that the Riemann structure is understood, a Riemann invariant boundary condition can be defined. The invariants are 
        \begin{equation}
            R_- = P_i + \frac{1}{4}(u^\perp_i)^2 + \zeta, \quad \mathrm{and} \quad R_+ = P_e + \frac{1}{4}(u^\perp_e)^2 + \zeta,
        \end{equation}
        where $i$ is the interior state and $e$ is the exterior state. The invariants clearly show that the system cannot support a solution with two rarefactions. Instead to apply a boundary condition a full solve for the Riemann problem is performed to get $u_b^\perp$ and $P_b$ based on the interior left state $u_i^\perp$, $P_i$, and the exterior right state $u_e^\perp$, $P_e$. The velocity at the boundary is then:
        \begin{equation}
            \mathbf{V}_b = \begin{cases}
                \mathbf{V}_i + \hat{\mathbf{n}}(u^\perp_b - \hat{\mathbf{n}}\cdot\mathbf{V}_i) &\mbox{if } \hat{\mathbf{n}}\cdot\mathbf{V}_i,\\
                \mathbf{V}_e + \hat{\mathbf{n}}(u^\perp_b - \hat{\mathbf{n}}\cdot\mathbf{V}_e) &\mbox{otherwise}.
            \end{cases}
        \end{equation}
        \newpage
\section{Additional Cylinder Plots}\label{app:cylinder_plots}
    
    \begin{figure}[tbhp]
        \centering
        \subfloat[ACM, $\zeta=3$, $\hat{t}=201$.]{\adjustbox{width=0.3\linewidth,valign=b}{\includegraphics[]{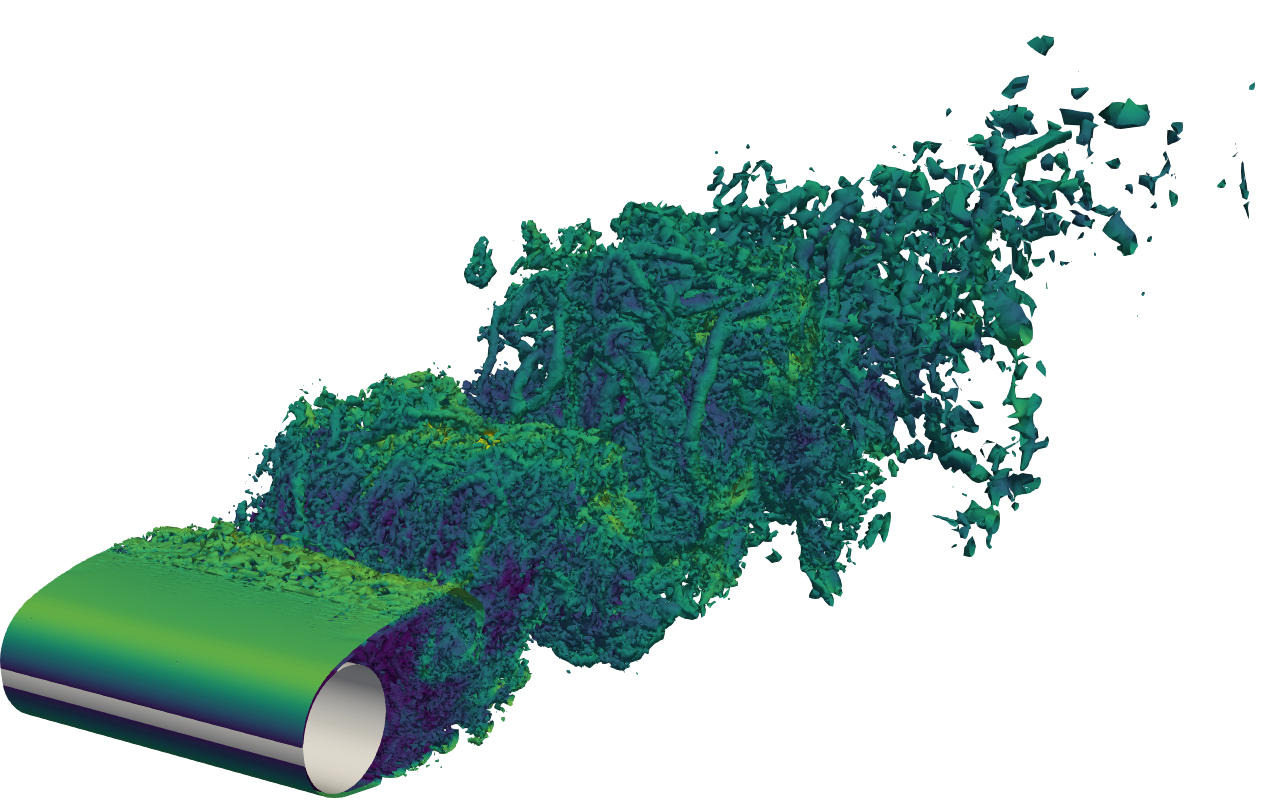}}}
        ~
        \subfloat[ACM, $\zeta=4.5$, $\hat{t}=201$.]{\adjustbox{width=0.3\linewidth,valign=b}{\includegraphics[]{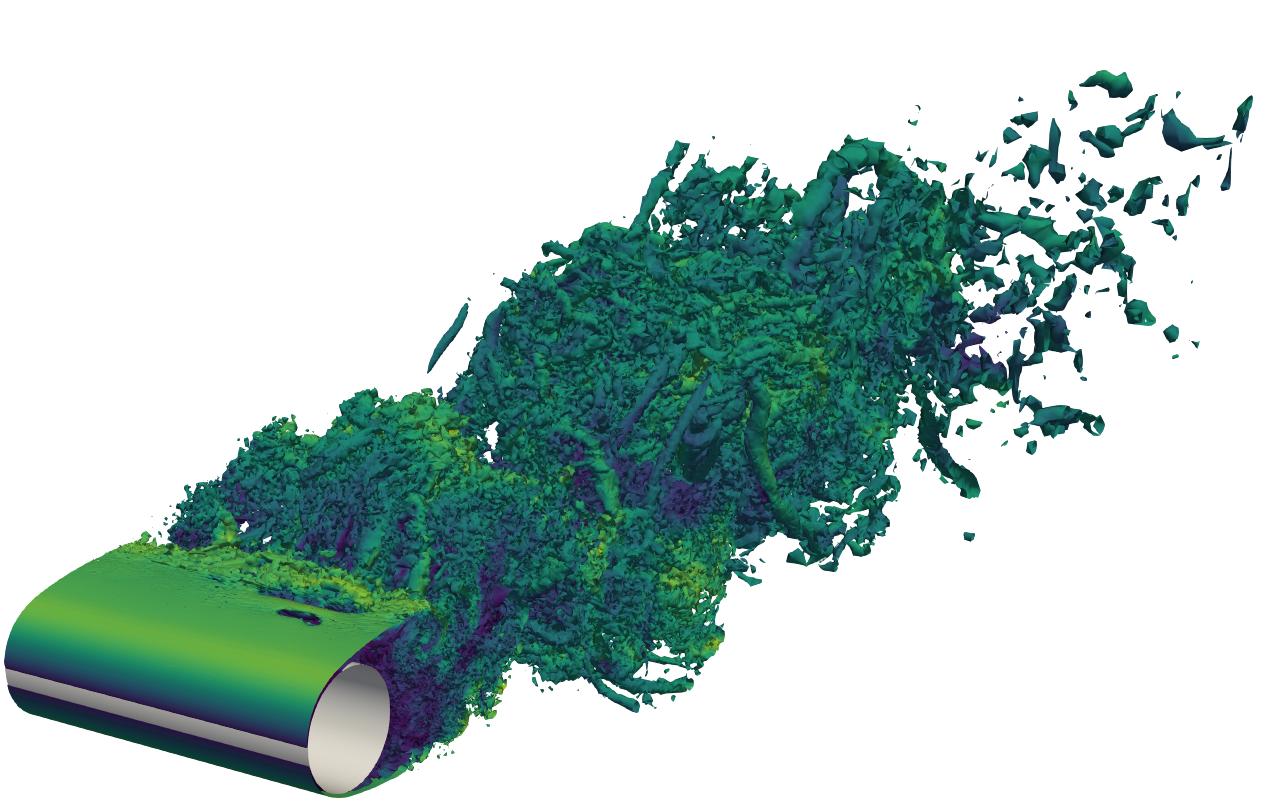}}}
        ~
        \subfloat[ACM-HD, $\zeta=3$, $T=100\nu$, $\hat{t}=201$.]{\adjustbox{width=0.3\linewidth,valign=b}{\includegraphics[]{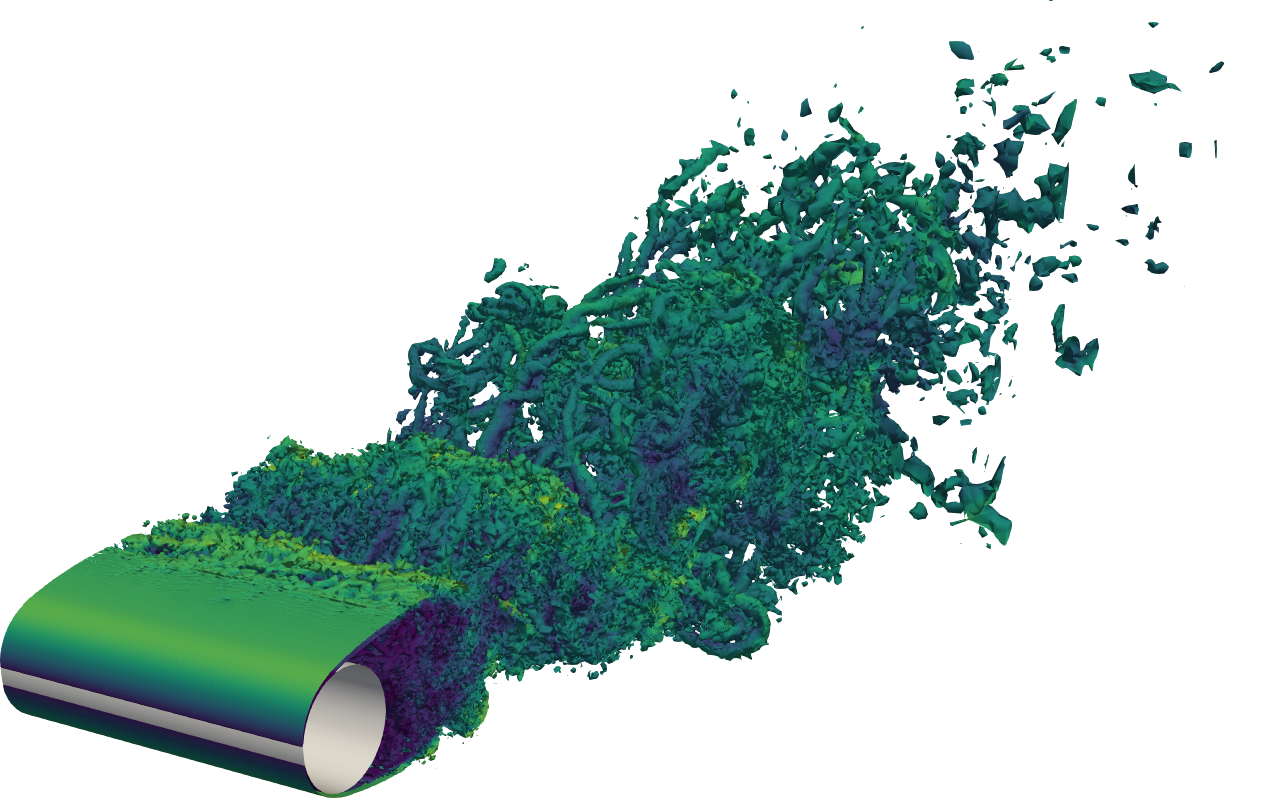}}}
        \\
        \subfloat[DNS, $\hat{t}=350$.]{\adjustbox{width=0.3\linewidth,valign=b}{\includegraphics[]{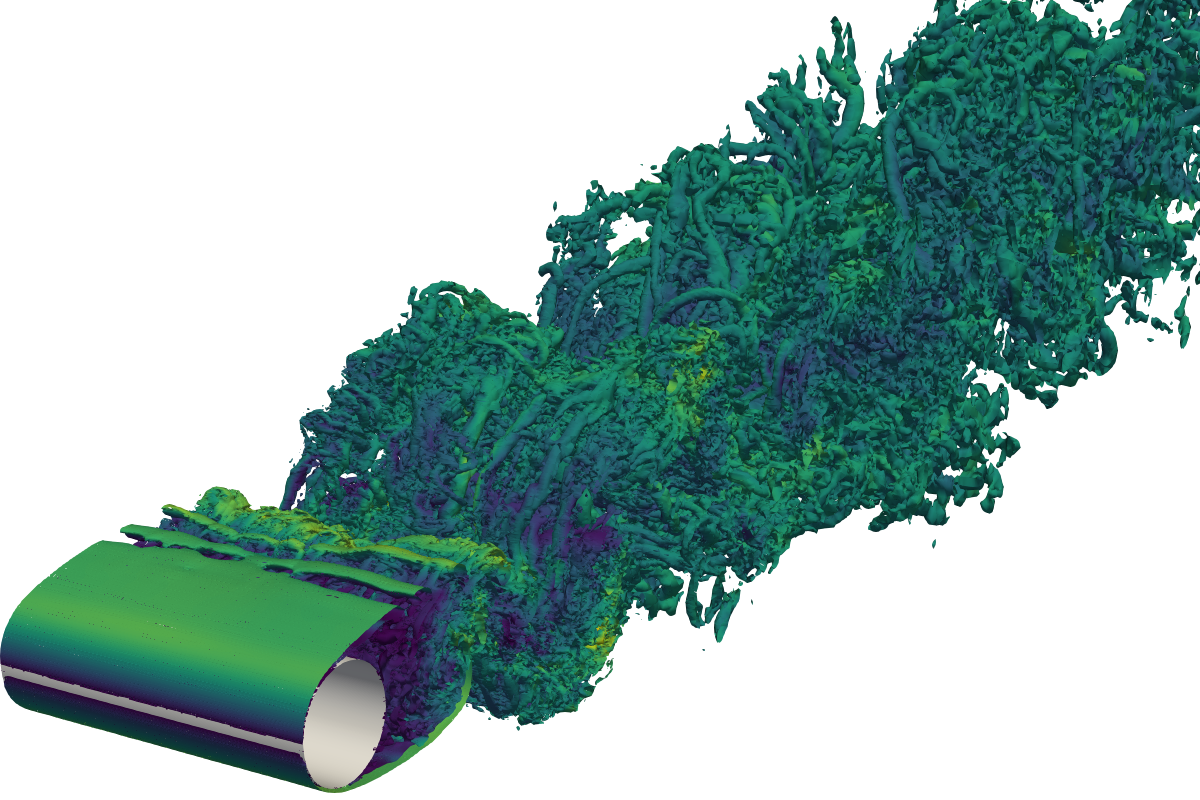}}}
        \caption{\label{fig:cylinder_q_extra}Iso-contours of Q-criterion, $Q=0.1$, coloured by velocity magnitude, DNS data courtesy of \citet{Dzanic2021}.}
    \end{figure}
    
    \begin{figure}[tbhp]
        \centering
        \subfloat[$\avg{u}/U$]{\adjustbox{width=0.47\linewidth, valign=b}{\input{figs/cylinder_u_centre}}}
        ~
        \subfloat[$\avg{u^\prime u^\prime}/U^2$]{\adjustbox{width=0.47\linewidth, valign=b}{\input{figs/cylinder_upup_centre}}}
        \caption{\label{fig:cylinder_centre}Centreline average quantities, with experimental data from \citet{Parnaudeau2008} and DNS data from \citet{Lehmkuhl2013}.}
    \end{figure}
    
    \begin{figure}[tbhp]
        \centering
        \subfloat[$\avg{u}/U$.]{\label{fig:cyl_u_1p54}\adjustbox{width=0.42\linewidth,valign=b}{\input{figs/cylinder_edac_u_y_x1-54}}}
        ~
        \subfloat[$\avg{v}/U$.]{\label{fig:cyl_v_1p54}\adjustbox{width=0.47\linewidth,valign=b}{\input{figs/cylinder_edac_v_y_x1-54}}}
        \\
        \subfloat[$\avg{u^\prime u^\prime}/U$.]{\label{fig:cyl_upup_1p54}\adjustbox{width=0.42\linewidth,valign=b}{\input{figs/cylinder_edac_upup_y_x1-54}}}
        ~
        \subfloat[$\avg{v^\prime v^\prime}/U$.]{\label{fig:cyl_vpvp_1p54}\adjustbox{width=0.42\linewidth,valign=b}{\input{figs/cylinder_edac_vpvp_y_x1-54}}}
        \caption{\label{fig:cylinder_x1-54}$x/D=1.54$, with experimental data from \citet{Parnaudeau2008} and DNS data from \citet{Lehmkuhl2013}.}
    \end{figure}

    \begin{figure}[tbhp]
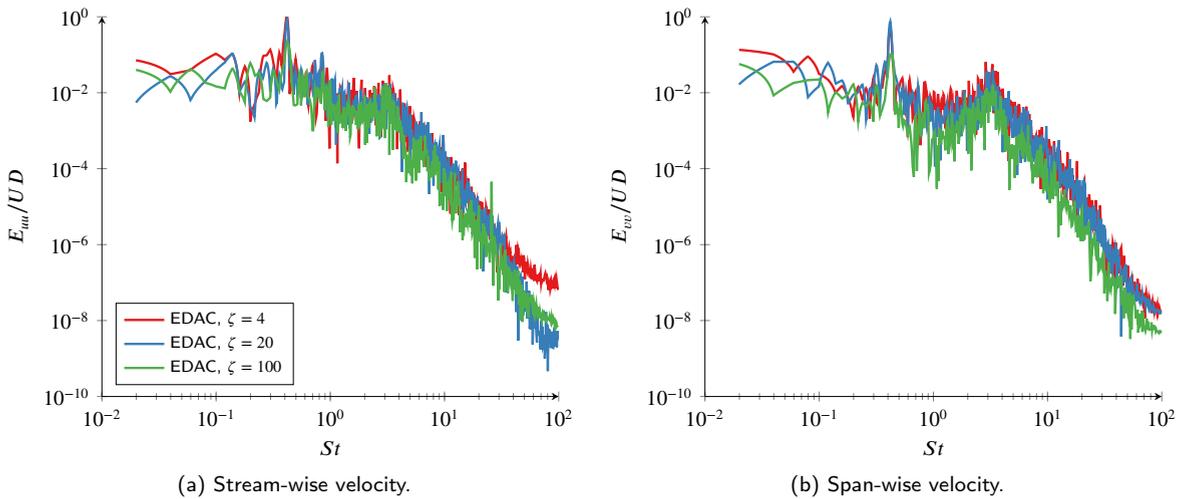

        \centering
        \subfloat[Stream-wise velocity.]{\adjustbox{width=0.47\linewidth, valign=b}{\input{figs/cylinder_spectra_u_Pnt1}}}
        ~
        \subfloat[Span-wise velocity.]{\adjustbox{width=0.47\linewidth, valign=b}{\input{figs/cylinder_spectra_v_Pnt1}}}
        \caption{\label{fig:cyl_psd}Power spectral density at $\mathbf{x} = (0.71D, 0.66D, 1.57D)$.}
    \end{figure}
    
\end{appendices}


\end{document}